\documentclass[aps,prl,reprint,superscriptaddress]{revtex4-1}
\usepackage{graphicx}
\usepackage[rightcaption]{sidecap}
\usepackage{braket}
\usepackage[utf8]{inputenc}
\usepackage{amsmath} 
\usepackage{amssymb}  
\usepackage{amstext}  

\begin{document}
\title{Phase-tuned entangled state generation between distant spin qubits}

\author{R. Stockill}
\altaffiliation{These authors contributed equally to the work.}
\affiliation{Cavendish Laboratory, University of Cambridge, JJ Thomson Avenue, Cambridge CB3 0HE, United Kingdom}

\author{M. J. Stanley}
\altaffiliation{These authors contributed equally to the work.}
\affiliation{Cavendish Laboratory, University of Cambridge, JJ Thomson Avenue, Cambridge CB3 0HE, United Kingdom}

\author{L. Huthmacher}
\altaffiliation{These authors contributed equally to the work.}
\affiliation{Cavendish Laboratory, University of Cambridge, JJ Thomson Avenue, Cambridge CB3 0HE, United Kingdom}

\author{E. Clarke}
\affiliation{EPSRC National Centre for III-V Technologies, University of Sheffield, Sheffield, S1 3JD, UK}

\author{M. Hugues}
\affiliation{Universit\'e C\^ote d'Azur, CNRS, CRHEA, Valbonne, France}

\author{A. J. Miller}
\affiliation{Quantum Opus, LLC, 45211 Helm St., Plymouth, MI 48170, USA}

\author{C. Matthiesen}
\altaffiliation[present address: ]{Department of Physics, University of California, Berkeley, California 94720, USA}
\affiliation{Cavendish Laboratory, University of Cambridge, JJ Thomson Avenue, Cambridge CB3 0HE, United Kingdom}

\author{C. Le Gall}
\affiliation{Cavendish Laboratory, University of Cambridge, JJ Thomson Avenue, Cambridge CB3 0HE, United Kingdom}

\author{M. Atat{\"u}re}
\email[Electronic address: ]{ma424@cam.ac.uk}
\affiliation{Cavendish Laboratory, University of Cambridge, JJ Thomson Avenue, Cambridge CB3 0HE, United Kingdom}

\begin{abstract}
Quantum entanglement between distant qubits is an important feature of quantum networks. Distribution of entanglement over long distances can be enabled through coherently interfacing qubit pairs via photonic channels. Here, we report the realization of optically generated quantum entanglement between electron spin qubits confined in two distant semiconductor quantum dots. The protocol relies on spin-photon entanglement in the trionic $\Lambda$-system and quantum erasure of the Raman-photon path. The measurement of a single Raman photon is used to project the spin qubits into a joint quantum state with an interferometrically stabilized and tunable relative phase. We report an average Bell-state fidelity for $\ket{\psi^{(+)}}$ and $\ket{\psi^{(-)}}$ states of $61.6\pm2.3\%$ and a record-high entanglement generation rate of 7.3 kHz between distant qubits. 
\end{abstract}


\date{\today}

\maketitle
The refutation of local realism in favor of a nonlocal quantum theory \cite{Einstein1935, Bell1964} has been supported by a number of decisive experiments, using entangled pairs of photons \cite{Freedman1972,Aspect1982}, atoms \cite{Rowe2001, Matsukevich2008} and solid-state systems \cite{Ansmann2009, Hensen2015}. While photonic links are essential to close loopholes in Bell tests by removing the requirement of spatial proximity for entanglement creation, they also permit flexible arrangements in which distant systems with a spin-photon interface can be entangled. The emergence of entanglement as a central resource in quantum sensing, communication and computing \cite{Wootters1998} benefits from this flexibility, where matter qubits coherently coupled to well-defined optical modes provide the elementary constituents of a distributed quantum network \cite{Kimble2008}. The operation rate of such a network would ultimately depend upon the oscillator strength of the spin-photon interface. In this regard, indium-gallium-arsenide (InGaAs) quantum dots (QDs) feature particularly strong light-matter coupling with the potential for high rates of entanglement distribution \cite{Lodahl2015}. Matter qubits can be realized using confined electrons \cite{Press2008}, heavy holes \cite{Brunner2009} or dark excitons \cite{Schwartz2015} in these systems. While each candidate qubit presents specific advantages \cite{Sun2016,Schwartz2016}, the electron spin has shown the longest coherence time thus far \cite{Press2010,Bechtold2015,Stockill2016}.

\indent In this Letter, we present optical generation of nonlocal quantum-entangled states between two distant nodes formed by electron spins confined in separate QDs. Through a single-photon state projection protocol \cite{Cabrillo1998} and the bright, narrow-linewidth emission available from QDs \cite{Kuhlmann2015} we realize an entanglement generation rate of 7.3 kHz, the highest rate to-date. Further, with full control over the single-photon interference, we create remote entangled states with arbitrary phase. Prior to this point, phase control of the generated entangled state had only been demonstrated for atomic nodes located in the same trap \cite{Slodicka2013}. Together with local gate operation times of a few picoseconds \cite{Press2008}, microsecond-long spin coherence \cite{Press2010,Bechtold2015,Stockill2016} and the current state-of-the-art for on-chip integration \cite{Lodahl2015, Luxmoore2013}, this work represents progress towards small-scale, on-chip quantum networks with high bandwidth operation.\\

\begin{figure}
\includegraphics[width=1\columnwidth,angle=0]{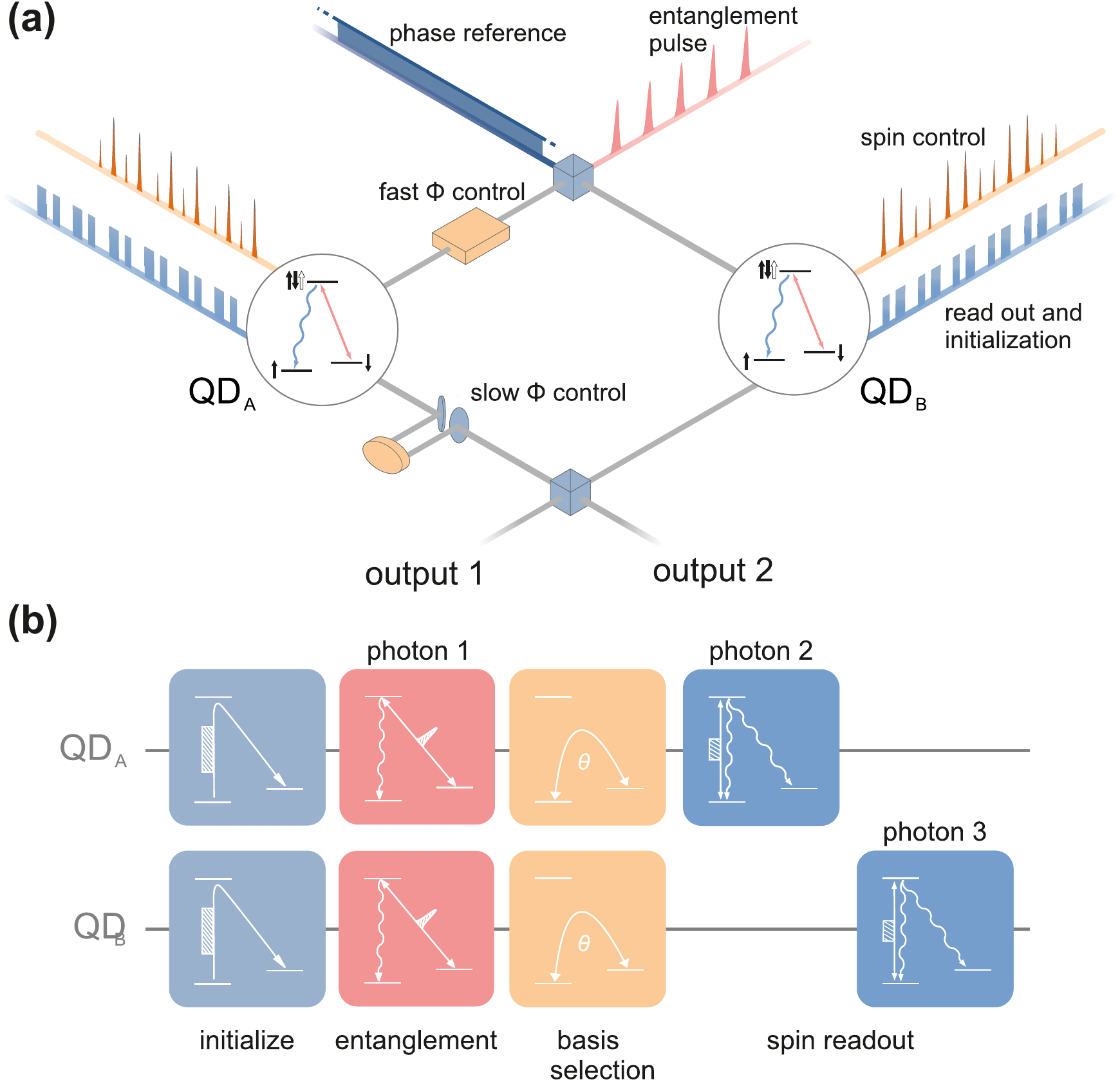}
\caption{\label{fig1} Experimental setup and sequence for detecting entanglement between distant spin qubits. (a) Sketch of the optical interferometer with the QD nodes. Three sets of optical pulses, labeled entanglement, read-out and rotation, are used in the sequence together with a continuous-wave laser for interferometer stabilization. Active phase stabilization to $<\pm3$ degrees is achieved using the combination of a retroreflector on a piezo-stack and a phase-electro-optic modulator. Spectral filtering (not illustrated here) after the output beam splitter allows separate detection of red and blue photons on superconducting nanowire single photon detectors, and the stabilizer signal on silicon single-photon counting modules. (b) Entanglement verification sequence. The spins are initialized in the $\left|\downarrow_{\mathrm{A}}\downarrow_{\mathrm{B}}\right\rangle$ state. The entanglement creation pulse is followed by $\pi/2$ rotations of both qubits when measuring in a transverse basis. Further $\pi$ rotations of the qubits are added to cycle through the four two-spin combinations in a given basis. Resonant excitation of the blue transition is used for read-out.}
\end{figure}

\indent The entanglement scheme employed in this work follows the proposal by Cabrillo et al. \cite{Cabrillo1998}, which relies on the phase-coherent excitation of two indistinguishable $\Lambda$-systems and the subsequent detection of a single state-projecting photon. First, a probabilistic Raman scattering process with amplitude $\sqrt{p}$ entangles the ground state of each system (spin-1/2 states $\ket{\uparrow_{i}}$ and $\ket{\downarrow_{i}}$) with the occupation of the Raman-photon mode \cite{Gao2012,Degreve2012,Schaibley2013}. Combining the Raman modes from each scatterer on a 50:50 beam splitter erases the which-path information, such that the total entangled spin-photon state is given by:\\
\begin{eqnarray}
\left|\Psi\right\rangle=&&(1-p)\left|\downarrow_{\mathrm{A}}\downarrow_\mathrm{B}\right\rangle\left|0_10_2\right\rangle \nonumber \\
&+&\sqrt{p(1-p)/2}\left(e^{i\phi_{\mathrm{A}}}\left|\uparrow_{\mathrm{A}}\downarrow_\mathrm{B}\right\rangle+e^{i\phi_\mathrm{B}}\left|\downarrow_{\mathrm{A}}\uparrow_\mathrm{B}\right\rangle\right)\left|1_10_2\right\rangle  \nonumber\\
&+&\sqrt{p(1-p)/2}\left(e^{i\phi_{\mathrm{A}}}\left|\uparrow_{\mathrm{A}}\downarrow_\mathrm{B}\right\rangle-e^{i\phi_\mathrm{B}}\left|\downarrow_{\mathrm{A}}\uparrow_\mathrm{B}\right\rangle\right)\left|0_11_2\right\rangle  \nonumber \\
&+&p/\sqrt{2} e^{i(\phi_{\mathrm{A}}+\phi_\mathrm{B})}\left|\uparrow_{\mathrm{A}}\uparrow_\mathrm{B}\right\rangle\left(\left|2_10_2\right\rangle-\left|0_12_2\right\rangle\right).
\end{eqnarray}
Here, the indices in the photonic number states designate the beam splitter output mode (1 or 2). $\phi_{\mathrm{A}}$ ($\phi_{\mathrm{B}}$) is the optical phase accumulated along the path going through system A (system B), transferred to the spin states through the beam splitter transformation. The state is sorted according to zero, one and two-photon contributions in the Raman mode.\\
 
\indent Detection of the one-photon contribution after the beam splitter projects the qubits into the maximally entangled state $\left(\ket{\uparrow_{\mathrm{A}}\downarrow_{\mathrm{B}}}\pm e^{i\Delta\phi}\ket{\downarrow_{\mathrm{A}}\uparrow_{\mathrm{B}}}\right)/\sqrt{2}$ with the sign depending on the output port that registers the Raman photon and $\Delta\phi = \phi_{\mathrm{B}} - \phi_{\mathrm{A}}$. For $\Delta\phi= 0$ these states correspond to the Bell states $\ket{\psi^{(\pm)}}$. With a probability of $p^2$ both qubits undergo a spin-flip scattering process, resulting in a two-photon state after the beam splitter. Without number-resolving detectors, and in the presence of optical losses, we cannot distinguish this component from the single-photon states, setting an intrinsic error in the Bell state generation of $p$. A compromise has to be found between this error and the probabilistic entangled-state generation rate, $2p(1-p)$. \\

\indent The experimental realization of this protocol is introduced in Fig. 1. Single electrons confined in two InGaAs QDs, QD$_{\mathrm{A}}$ and QD$_{\mathrm{B}}$, located approx. 2 meters apart in separate cryostats provide the stationary nodes. $4$-T magnetic field applied perpendicular to the QD growth axis lifts the spin-degeneracy of the ground and excited states within each node.
The 25-GHz-split electron spin ground states, together with one of the two excited states form a $\Lambda$-system [Fig. 1(a)]. The two optical transitions, at 968 nm in our case, are distinguished by their frequency: the lower-energy and higher-energy transitions are denoted red and blue, respectively. The entangled state phase, $\Delta\phi$, is defined by the general architecture of our setup: a Mach-Zehnder Interferometer (MZI) where the mirrors in the two arms are replaced by the two QDs [Fig. 1(a)]. \\

\indent The measurement sequence is illustrated in Fig. 1(b). A cycle starts with optical pumping of both spins into the spin down state ($\approx97\%$ fidelity). We then apply a 160-ps long pulse to the red transition of each QD to generate the state-projecting Raman photon. The state-change probability, $p$, is set to 7\% in order to suppress the aforementioned error in the entangled state due to simultaneous spin-flip Raman emission events. Having detected a state-projecting Raman photon [photon $1$ in Fig. 1(b)], evidence of entanglement between the spins is obtained from spin correlations in different measurement bases [photons $2$ and $3$ in Fig. 1(b)]. 
We combine local spin-rotation with read-out of the spin-up state to reconstruct the population of $\{\ket{\downarrow_{\mathrm{A}} \downarrow_{\mathrm{B}}}$, $\ket{\downarrow_{\mathrm{A}} \uparrow_{\mathrm{B}}}$, $\ket{\uparrow_{\mathrm{A}} \downarrow_{\mathrm{B}}}$, $\ket{\uparrow_{\mathrm{A}} \uparrow_{\mathrm{B}}}\}$ in four iterations of the sequence. The read-out of each QD is performed in turn with 8-ns long pulses resonant with the blue transition. An optional $\pi$/2 rotation of both electron spins allows correlations to be measured in a transverse basis. These rotations occur within the inhomogeneous dephasing time ($T_2^*$) of $1.2$ ns after the entanglement pulse during which the phase of the two-spin state is preserved \cite{Merkulov2002}. \\

\indent Spin initialization, entanglement and read-out pulses are derived from frequency-stabilized single-mode lasers using fiber-based electro-optic intensity modulators. Local spin rotations are performed optically \cite{Press2008} using $1$-THz red-detuned pulses picked from a mode-locked Ti:Sa laser with acousto-optic modulators. The whole experiment is locked to a clock set by the Ti:Sa laser, with a sequence length of  78.9 ns and a repetition rate of 10.9 MHz \cite{supplementary}. We use a far-off resonance laser to monitor the phase of the MZI continuously. The combination of electro-optic phase modulation and piezo-based compensation [fast and slow $\phi$ control in Fig. 1(a)] allows us to actively stabilize to $\le$$\pm3$ degrees over a DC$-1.5$ kHz range. The working point of the MZI  and the timing of the spin rotations are arranged such that we can determine \textit{a priori} the phase of the entangled state we create by monitoring the interference of coherent Rayleigh scattering from the two emitters \cite{supplementary}. After the output beam splitter, gratings and Fabry-P\'{e}rot filters are used to separate the phase-reference laser, and the red and blue QD fluorescence onto 6 single-photon detectors. A time-to-digital converter time-tags photon detection events originating from the QDs for analysis \cite{supplementary}.\\

\begin{figure}
\includegraphics[width=1\columnwidth,angle=0]{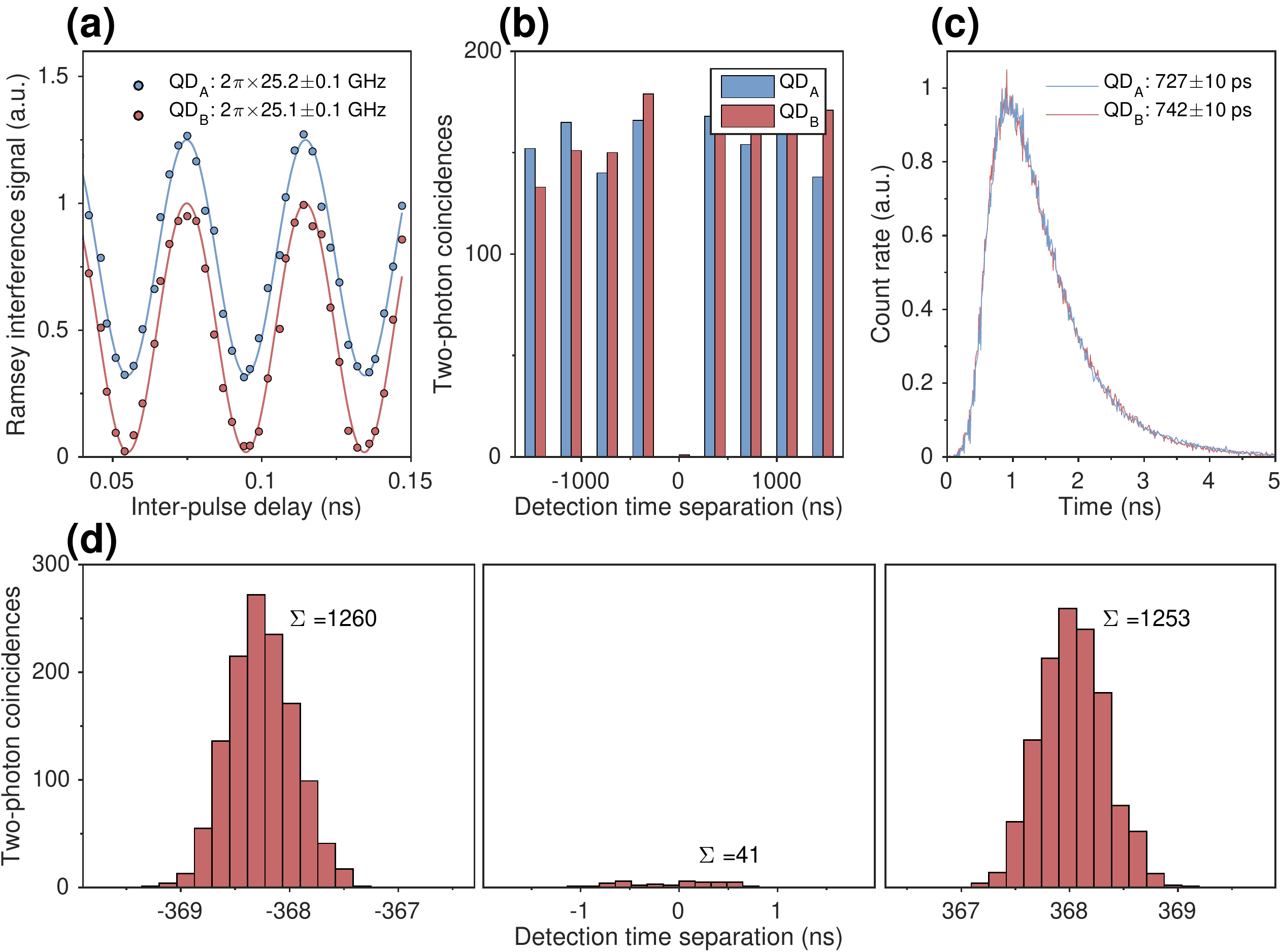}
\caption{\label{fig2}  Overlap of the optical and spin properties (a) Ramsey interference fringes for both QDs. The curves have been offset for clarity. Sinusoidal fits to the data show electron spins splittings around $2\pi\times25.1$ GHz. This splitting is inhomogeneously broadened by the hyperfine interaction (FWHM $\approx2\pi\times100$ MHz) (b) Two-photon correlation measurements of the Raman emission generated by the entanglement pulse for each QD. In the course of three minutes of data acquisition, we register no zero delay coincidence events for QD$_{\mathrm{A}}$ and a single event for QD$_{\mathrm{B}}$, consistent with the limit set by our laser-background count rate. (c) Radiative decay measurements of the Raman transitions. (d) Raman two photon indistinguishability measurement with the same count rate from each QD. The three peaks closest to zero-time delay are displayed. The measured Hong-Ou-Mandel indistinguishability is $93\pm1\%$.}
\end{figure}

\indent Projecting a well-defined entangled state with a stationary relative phase requires identical $\Lambda$-systems for the two QDs \cite{Vittorini2014}. First, we evaluate the ground state energy splitting using Ramsey interferometry [Fig. 2(a)] and adjust the Zeeman energy of the spin qubits by fine-tuning the external magnetic fields around 4 T. The spin precession frequencies for the two QDs are closely matched at around $2\pi\times25.1$ GHz, ensuring that the phase $\Delta\phi$ is stationary in the laboratory frame. Under these conditions the Zeeman energy far exceeds the optical linewidth and each QD acts as a photon turnstile, whereby a Raman scattering process shelves the electron spin, preventing subsequent excitation. Figure 2(b) displays the second-order autocorrelation measurements of the Raman scattering from each QD following the entanglement pulse, obtained by blocking the MZI arms in turn.  The anti-bunching is limited in our case by laser background and detector dark counts, which contribute 1 in 150 events on average, consistent with the single two-photon event recorded at zero time delay. In parallel we must consider the optical mode-matching between the two QDs. Figure 2(c) shows very similar excited state lifetimes: $727 \pm 10$ ps ($742 \pm 10$ ps) for QD$_{\mathrm{A}}$ (QD$_{\mathrm{B}}$), which guarantees photon-wavepacket overlap in time, while a static electric field applied across each sample is used to overlap the photons spectrally via the Stark effect. Our setup filters the phonon-assisted emission ($\approx 10\%$) that occurs outside of a well-defined frequency mode. We implement a Hong-Ou-Mandel experiment \cite{Hong1987} to quantify the quality of the quantum erasure process by cross-correlating the Raman photons in the two beam splitter outputs following the entanglement pulse. The central three peaks of the correlation are shown in Fig. 2(d), formed of coincidences found in the same or consecutive sequence repetitions. The small number of coincidences in the central peak is the signature of photon coalescence owing to indistinguishability of the input modes, revealing a two-photon interference visibility of $93 \pm 1\%$. Imperfect optical elements account for a $4\%$-reduction of the visibility in our setup. We only consider photons within a $1.2$-ns window from the start of the entanglement pulse, limiting the effect of inhomogeneous spin dephasing on Raman photon distinguishability \cite{Santori2009}, with the condition that we reject 42\% of the emission events.\\

\begin{figure*}
\centering
\includegraphics[trim = 0mm 0mm 0mm 0mm, clip, width=.94\textwidth]{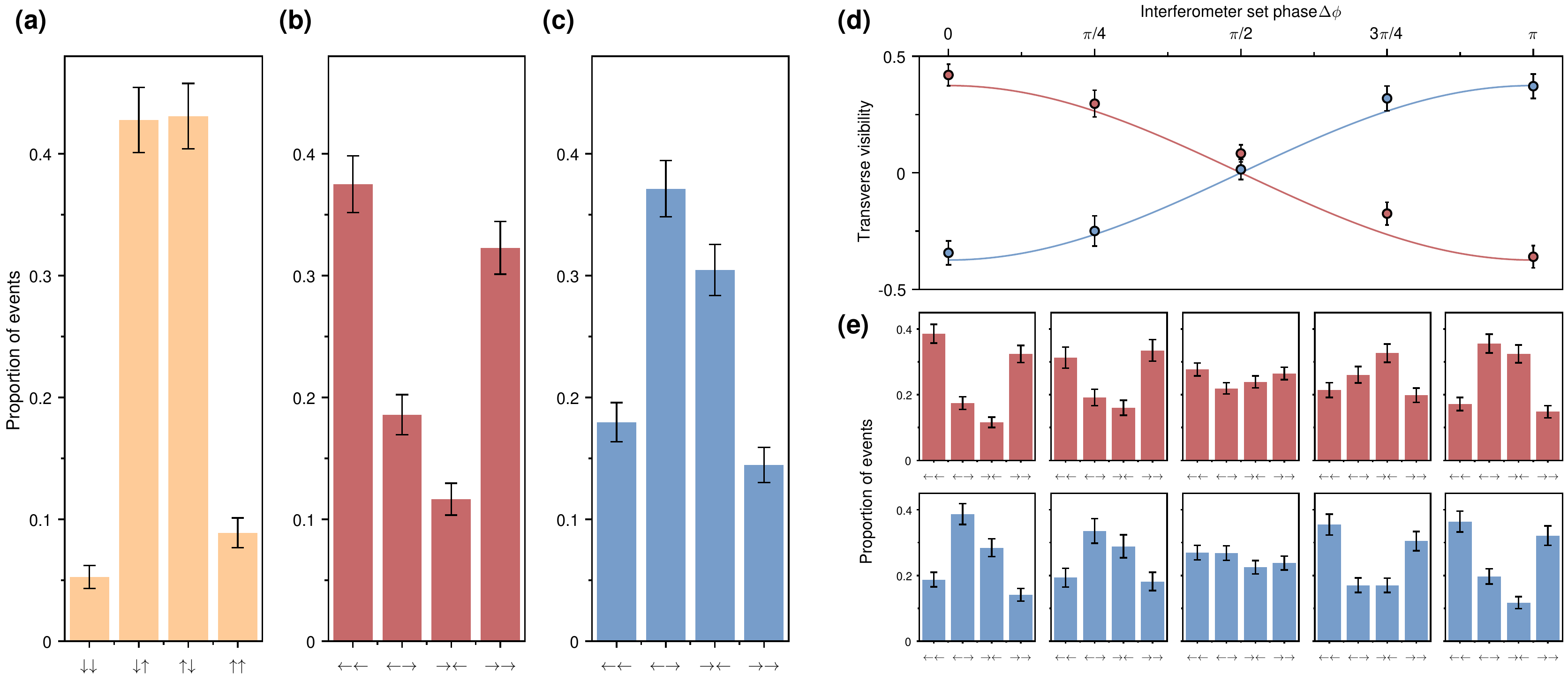}
\caption{\label{fig3} Joint spin state reconstruction through 3-photon coincidence events. (a) Joint spin state population conditional on a single Raman photon detection event. An antisymmetric population is retrieved with a fidelity of $85.7 \pm 3.8\%$. The error bars represent the statistical uncertainty drawn from the $603$ three-photon events that were used to reconstruct the population.  (b-c) Spin correlations in the transverse basis. (d-e) Phase control in the entanglement generation step. (d) Visibility in the transverse basis for varying optical phase between the interfering paths. A positive visibility corresponds to a correlated phase, and negative to an anti-correlated phase. The curves are sinusoidal fits to the data. (e) Normalized coincidence events for a Raman photon detected on beam splitter output 1 (red histograms) or output 2 (blue histograms) for five different set points of the interferometer phase.}
\end{figure*}

\indent In Fig. 3 we reconstruct the projected two-spin state from the correlation of three-photon coincidence events. We first measure the state in the population basis, parallel to the external magnetic field [Fig. 3 (a)]. We find that the detection of a Raman photon predicts an antisymmetric spin population with a probability of $85.7\pm3.8\%$. The uncertainty is set by the shot noise of the $603$ three-photon events that contribute to the data in Fig. 3(a). The presence of population in the $\ket{\uparrow_{\mathrm{A}} \uparrow_{\mathrm{B}}}$ state is intrinsic to the entanglement generation scheme, and follows the spin-flip probability $p$. Events in the $\ket{\downarrow_{\mathrm{A}} \downarrow_{\mathrm{B}}}$ state mainly result from imperfect spin rotation prior to the population read-out \cite{supplementary} as well as read-out laser leakage. Additional rotations in our pulse sequence [Fig. 1(b)] transfer the phase of the entangled state into correlated spin populations. We stabilize the Raman mode at either $\Delta\phi=0$ or $\Delta\phi=\pi$ to obtain maximum contrast in our state tomography. Figure 3(b-c) shows correlation measurements in this transverse basis. For $\Delta\phi=0$, Raman-photon detection in output mode 1 generates the $\ket{\psi^{(+)}}$ state while detection in mode 2 generates $\ket{\psi^{(-)}}$. When the Raman mode is stabilized at $\Delta\phi=\pi$, output mode 1(2) heralds the creation of $\ket{\psi^{(-)}}$($\ket{\psi^{(+)}}$). In Fig. 3(b-c) the coincidence events are displayed accordingly, with 3(b) corresponding to projection of $\ket{\psi^{(+)}}$ and 3(c) to $\ket{\psi^{(-)}}$.\\

\indent In the rotated basis, the $\pi$ phase-shift between the states generated at opposite output ports is apparent from the correlation histograms, revealing visibilities of $35.1 \pm 3.8\%$ for $\ket{\psi^{(+)}}$ [Fig. 3(b)] and $39.5 \pm 3.8\%$ for $\ket{\psi^{(-)}}$ [Fig. 3(c)]. Owing to the rapid phase evolution of all but a subset of spin coherences in the density matrix \cite{supplementary}, we can directly estimate the Bell-state fidelity from these visibilities. Overall, we find that a photon detection heralds either the $\ket{\psi^{(+)}}$ or $\ket{\psi^{(-)}}$ state with an average of $61.6\pm2.3\%$ fidelity. For these results the non-classicality of the state is confirmed to over 5 standard deviations of the mean. The extracted fidelity can be understood by taking into account the contributions of the double spin flip rate (limiting the fidelity to 93\%), the imperfect mode overlap (a further reduction of 4\%), the spin-state dephasing (13\%), and the imperfect spin preparation and read-out (3\% and 6\% respectively). The combination of these factors predict a fidelity of 71\%. The most likely source for our observed reduction below this value is electrical noise in each QD sample \cite{Kuhlmann2013,Matthiesen2014} which alters the relative phase of the two optical excitations, and in turn, the phase of the entangled state \cite{supplementary}. \\

\indent In Fig. 3(d-e) we demonstrate direct control over the generated two-electron spin state. Figure 3(d) shows the extracted visibilities from transverse basis measurements for five different values of the entangled-state phase. The visibilities are drawn from the relative coincidence rates displayed in Fig. 3(e). We observe a sinusoidal variation in the visibility of the phase-basis correlations as the MZI phase $\Delta\phi$ is tuned from $0$ to $\pi$. The results are partitioned according to the interferometer output that registers the Raman photon. For  $\Delta\phi= 0$  the first (second) output projects the spins to $\ket{\psi^{(+)}}$ ($\ket{\psi^{(-)}}$), resulting in correlated (anti-correlated) population in the transverse measurement. As we move to $\Delta\phi=\pi/2$, we project to $\ket{\psi^{(+)}\pm\psi^{(-)}}$ which do not exhibit transverse spin-spin correlations. At $\Delta\phi=\pi$, the relationships between the detection modes and the spin states are swapped and the visibility along our measurement axis returns. This state control demonstrates the working principle of our experiment: quantum erasure imprints the interferometric phase between the Raman modes onto the nonlocal state shared between the two spins.\\

\begin{figure}
\includegraphics[trim = 0mm 0mm 0mm 0mm, clip, width=1\columnwidth,angle=0]{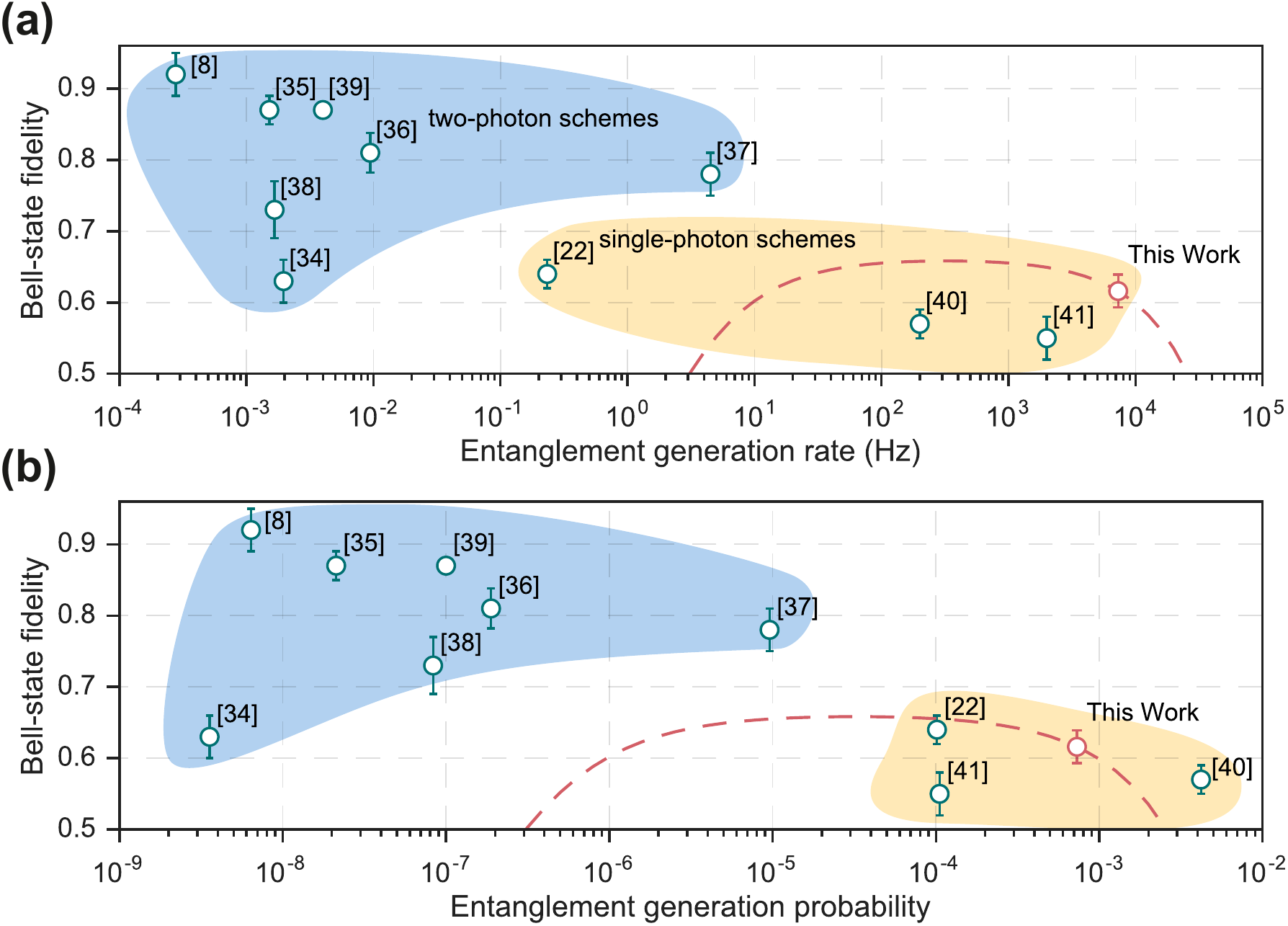}
\caption{\label{fig4} Comparison of the extracted fidelity and entanglement generation rate (a) and generation probability (b) between the electron spins in this work and other physical systems. The indices denote the reference of the work the figures were drawn from. The shaded regions mark the distribution protocol used, either single or double photon detection state projection. We note our quoted fidelity is not corrected for our imperfect state retrieval.}
\end{figure}

\indent In Fig. 4 we compare our results with entanglement rates and state fidelities reported thus far for distant qubits using projection-by-measurement protocols. Figure 4(a) shows the measured Bell-state fidelities plotted against the entanglement generation rate for experiments in atomic qubits \cite{Moehring2007,Maunz2009,Hofmann2012,Slodicka2013,Hucul2014}, NV centers \cite{Bernien2013, Pfaff2014}, superconducting qubits \cite{Narla2016} and confined hole spins in QDs \cite{Delteil2015} along with the average fidelity for electron spins in our work. At $7.3$ kHz, we are reporting the fastest distribution of entanglement between distant qubits. Solid-state systems do not require intermittent cooling or reloading steps and as such our protocol can operate continually at a 10.9-MHz attempt rate. Further insight can be obtained from the success probability, $p_{\mathrm{succ}}$, which factors out the intrinsic timescales of each physical system and the specific requirements of each control sequence, which we plot in Fig. 4(b). The single-photon protocol we use allows for a success probability of $p_{\mathrm{succ}} = 6.7\times10^{-4}$, the highest reported for optical-frequency qubits. In comparison, two-photon schemes have been reported to operate with $p_{\mathrm{succ}}$ below $1\times10^{-5}$. In particular, our demonstration benefits from superconducting nanowire single photon detectors with a detection efficiency of over 80$\%$ at the operation wavelength. The success probability of the single-photon scheme is intimately linked to the Bell-state fidelity. To better understand this relationship, the dashed red curves in Fig. 4 project our state fidelity over different entanglement generation rates, establishing the reach of our experimental protocol. Reducing double spin-flip events would increase the fidelity at the expense of lower scattering probabilities, accounting for the intrinsic and technical limitations discussed earlier, until we reach the $\approx1$-Hz detector dark count rate where false heralds dominate.\\

\indent In this work, we demonstrate optically generated quantum entanglement between two confined electron spins. In order to use this probabilistic scheme in computational protocols, a herald is crucial to allow a repeat-until-success approach. Owing to the short ensemble dephasing time of the electron spins, we apply the spin read-out sequence before the Raman photon reaches the detector and heralds entanglement. Beyond this proof-of-principle demonstration, local decoupling operations would preserve the coherence of the joint quantum state beyond the required propagation time of the state-heralding photon ($\approx$100-ns in our system) \cite{Press2010,Bechtold2015,Stockill2016}. In order to extend to multiple nodes and realize fault-tolerant operation, we require both distribution rates within the storage time of the entangled state and additional memory qubits. Using current state-of-the-art light-collection strategies \cite{Somaschi2015,Wang2016}, the entanglement rate could be improved to $130$ kHz. This rate approaches the inverse of electron-spin coherence times in self-assembled QDs \cite{Press2010,Stockill2016,Prechtel2016}, a crucial benchmark for fault tolerant scalability \cite{Monroe2014}. Through further improvement of this rate, the second spin of a quantum dot molecule \cite{Kim2011} could provide a viable memory qubit. Alternatively, hybrid structures such as a capacitively coupled electrostatically defined quantum dot \cite{Kim2016} could reach the threshold through the increase of the state coherence time. \\

\indent We gratefully acknowledge financial support by the European Research Council ERC Consolidator grant agreement no. 617985 and the EPSRC
National Quantum Technologies Programme NQIT EP/M013243/1. C.M. acknowledges Clare College, Cambridge, for financial support through a Junior Research Fellowship. We thank A. Imamoglu and B. Munro for fruitful discussions and S. Topliss for technical assistance.


\begin{thebibliography}{47}%
\makeatletter
\providecommand \@ifxundefined [1]{%
 \@ifx{#1\undefined}
}%
\providecommand \@ifnum [1]{%
 \ifnum #1\expandafter \@firstoftwo
 \else \expandafter \@secondoftwo
 \fi
}%
\providecommand \@ifx [1]{%
 \ifx #1\expandafter \@firstoftwo
 \else \expandafter \@secondoftwo
 \fi
}%
\providecommand \natexlab [1]{#1}%
\providecommand \enquote  [1]{``#1''}%
\providecommand \bibnamefont  [1]{#1}%
\providecommand \bibfnamefont [1]{#1}%
\providecommand \citenamefont [1]{#1}%
\providecommand \href@noop [0]{\@secondoftwo}%
\providecommand \href [0]{\begingroup \@sanitize@url \@href}%
\providecommand \@href[1]{\@@startlink{#1}\@@href}%
\providecommand \@@href[1]{\endgroup#1\@@endlink}%
\providecommand \@sanitize@url [0]{\catcode `\\12\catcode `\$12\catcode
  `\&12\catcode `\#12\catcode `\^12\catcode `\_12\catcode `\%12\relax}%
\providecommand \@@startlink[1]{}%
\providecommand \@@endlink[0]{}%
\providecommand \url  [0]{\begingroup\@sanitize@url \@url }%
\providecommand \@url [1]{\endgroup\@href {#1}{\urlprefix }}%
\providecommand \urlprefix  [0]{URL }%
\providecommand \Eprint [0]{\href }%
\providecommand \doibase [0]{http://dx.doi.org/}%
\providecommand \selectlanguage [0]{\@gobble}%
\providecommand \bibinfo  [0]{\@secondoftwo}%
\providecommand \bibfield  [0]{\@secondoftwo}%
\providecommand \translation [1]{[#1]}%
\providecommand \BibitemOpen [0]{}%
\providecommand \bibitemStop [0]{}%
\providecommand \bibitemNoStop [0]{.\EOS\space}%
\providecommand \EOS [0]{\spacefactor3000\relax}%
\providecommand \BibitemShut  [1]{\csname bibitem#1\endcsname}%
\let\auto@bib@innerbib\@empty
\bibitem [{\citenamefont {Einstein}\ \emph {et~al.}(1935)\citenamefont
  {Einstein}, \citenamefont {Podolsky},\ and\ \citenamefont
  {Rosen}}]{Einstein1935}%
  \BibitemOpen
  \bibfield  {author} {\bibinfo {author} {\bibfnamefont {A.}~\bibnamefont
  {Einstein}}, \bibinfo {author} {\bibfnamefont {B.}~\bibnamefont {Podolsky}},
  \ and\ \bibinfo {author} {\bibfnamefont {N.}~\bibnamefont {Rosen}},\ }\href
  {\doibase 10.1103/PhysRev.47.777} {\bibfield  {journal} {\bibinfo  {journal}
  {Physical Review}\ }\textbf {\bibinfo {volume} {47}},\ \bibinfo {pages} {777}
  (\bibinfo {year} {1935})}\BibitemShut {NoStop}%
\bibitem [{\citenamefont {Bell}(1964)}]{Bell1964}%
  \BibitemOpen
  \bibfield  {author} {\bibinfo {author} {\bibfnamefont {J.~S.}\ \bibnamefont
  {Bell}},\ }\href@noop {} {\bibfield  {journal} {\bibinfo  {journal}
  {Physics}\ }\textbf {\bibinfo {volume} {1}},\ \bibinfo {pages} {195}
  (\bibinfo {year} {1964})}\BibitemShut {NoStop}%
\bibitem [{\citenamefont {Freedman}\ and\ \citenamefont
  {Clauser}(1972)}]{Freedman1972}%
  \BibitemOpen
  \bibfield  {author} {\bibinfo {author} {\bibfnamefont {S.~J.}\ \bibnamefont
  {Freedman}}\ and\ \bibinfo {author} {\bibfnamefont {J.~F.}\ \bibnamefont
  {Clauser}},\ }\href {\doibase 10.1103/PhysRevLett.28.938} {\bibfield
  {journal} {\bibinfo  {journal} {Physical Review Letters}\ }\textbf {\bibinfo
  {volume} {28}},\ \bibinfo {pages} {938} (\bibinfo {year} {1972})}\BibitemShut
  {NoStop}%
\bibitem [{\citenamefont {Aspect}\ \emph {et~al.}(1982)\citenamefont {Aspect},
  \citenamefont {Dalibard},\ and\ \citenamefont {Roger}}]{Aspect1982}%
  \BibitemOpen
  \bibfield  {author} {\bibinfo {author} {\bibfnamefont {A.}~\bibnamefont
  {Aspect}}, \bibinfo {author} {\bibfnamefont {J.}~\bibnamefont {Dalibard}}, \
  and\ \bibinfo {author} {\bibfnamefont {G.}~\bibnamefont {Roger}},\ }\href
  {\doibase 10.1103/PhysRevLett.49.1804} {\bibfield  {journal} {\bibinfo
  {journal} {Physical Review Letters}\ }\textbf {\bibinfo {volume} {49}},\
  \bibinfo {pages} {1804} (\bibinfo {year} {1982})}\BibitemShut {NoStop}%
\bibitem [{\citenamefont {Rowe}\ \emph {et~al.}(2001)\citenamefont {Rowe},
  \citenamefont {Kielpinski}, \citenamefont {Meyer}, \citenamefont {Sackett},
  \citenamefont {Itano}, \citenamefont {Monroe},\ and\ \citenamefont
  {Wineland}}]{Rowe2001}%
  \BibitemOpen
  \bibfield  {author} {\bibinfo {author} {\bibfnamefont {M.~A.}\ \bibnamefont
  {Rowe}}, \bibinfo {author} {\bibfnamefont {D.}~\bibnamefont {Kielpinski}},
  \bibinfo {author} {\bibfnamefont {V.}~\bibnamefont {Meyer}}, \bibinfo
  {author} {\bibfnamefont {C.~A.}\ \bibnamefont {Sackett}}, \bibinfo {author}
  {\bibfnamefont {W.~M.}\ \bibnamefont {Itano}}, \bibinfo {author}
  {\bibfnamefont {C.}~\bibnamefont {Monroe}}, \ and\ \bibinfo {author}
  {\bibfnamefont {D.~J.}\ \bibnamefont {Wineland}},\ }\href {\doibase
  10.1038/35057215} {\bibfield  {journal} {\bibinfo  {journal} {Nature}\
  }\textbf {\bibinfo {volume} {409}},\ \bibinfo {pages} {791} (\bibinfo {year}
  {2001})}\BibitemShut {NoStop}%
\bibitem [{\citenamefont {Matsukevich}\ \emph {et~al.}(2008)\citenamefont
  {Matsukevich}, \citenamefont {Maunz}, \citenamefont {Moehring}, \citenamefont
  {Olmschenk},\ and\ \citenamefont {Monroe}}]{Matsukevich2008}%
  \BibitemOpen
  \bibfield  {author} {\bibinfo {author} {\bibfnamefont {D.~N.}\ \bibnamefont
  {Matsukevich}}, \bibinfo {author} {\bibfnamefont {P.}~\bibnamefont {Maunz}},
  \bibinfo {author} {\bibfnamefont {D.~L.}\ \bibnamefont {Moehring}}, \bibinfo
  {author} {\bibfnamefont {S.}~\bibnamefont {Olmschenk}}, \ and\ \bibinfo
  {author} {\bibfnamefont {C.}~\bibnamefont {Monroe}},\ }\href {\doibase
  10.1103/PhysRevLett.100.150404} {\bibfield  {journal} {\bibinfo  {journal}
  {Physical Review Letters}\ }\textbf {\bibinfo {volume} {100}},\ \bibinfo
  {pages} {150404} (\bibinfo {year} {2008})}\BibitemShut {NoStop}%
\bibitem [{\citenamefont {Ansmann}\ \emph {et~al.}(2009)\citenamefont
  {Ansmann}, \citenamefont {Wang}, \citenamefont {Bialczak}, \citenamefont
  {Hofheinz}, \citenamefont {Lucero}, \citenamefont {Neeley}, \citenamefont
  {O'Connell}, \citenamefont {Sank}, \citenamefont {Weides}, \citenamefont
  {Wenner}, \citenamefont {Cleland},\ and\ \citenamefont
  {Martinis}}]{Ansmann2009}%
  \BibitemOpen
  \bibfield  {author} {\bibinfo {author} {\bibfnamefont {M.}~\bibnamefont
  {Ansmann}}, \bibinfo {author} {\bibfnamefont {H.}~\bibnamefont {Wang}},
  \bibinfo {author} {\bibfnamefont {R.~C.}\ \bibnamefont {Bialczak}}, \bibinfo
  {author} {\bibfnamefont {M.}~\bibnamefont {Hofheinz}}, \bibinfo {author}
  {\bibfnamefont {E.}~\bibnamefont {Lucero}}, \bibinfo {author} {\bibfnamefont
  {M.}~\bibnamefont {Neeley}}, \bibinfo {author} {\bibfnamefont {A.~D.}\
  \bibnamefont {O'Connell}}, \bibinfo {author} {\bibfnamefont {D.}~\bibnamefont
  {Sank}}, \bibinfo {author} {\bibfnamefont {M.}~\bibnamefont {Weides}},
  \bibinfo {author} {\bibfnamefont {J.}~\bibnamefont {Wenner}}, \bibinfo
  {author} {\bibfnamefont {A.~N.}\ \bibnamefont {Cleland}}, \ and\ \bibinfo
  {author} {\bibfnamefont {J.~M.}\ \bibnamefont {Martinis}},\ }\href {\doibase
  10.1038/nature08363} {\bibfield  {journal} {\bibinfo  {journal} {Nature}\
  }\textbf {\bibinfo {volume} {461}},\ \bibinfo {pages} {504} (\bibinfo {year}
  {2009})}\BibitemShut {NoStop}%
\bibitem [{\citenamefont {Hensen}\ \emph {et~al.}(2015)\citenamefont {Hensen},
  \citenamefont {Bernien}, \citenamefont {Dr{\'{e}}au}, \citenamefont
  {Reiserer}, \citenamefont {Kalb}, \citenamefont {Blok}, \citenamefont
  {Ruitenberg}, \citenamefont {Vermeulen}, \citenamefont {Schouten},
  \citenamefont {Abell{\'{a}}n}, \citenamefont {Amaya}, \citenamefont
  {Pruneri}, \citenamefont {Mitchell}, \citenamefont {Markham}, \citenamefont
  {Twitchen}, \citenamefont {Elkouss}, \citenamefont {Wehner}, \citenamefont
  {Taminiau},\ and\ \citenamefont {Hanson}}]{Hensen2015}%
  \BibitemOpen
  \bibfield  {author} {\bibinfo {author} {\bibfnamefont {B.}~\bibnamefont
  {Hensen}}, \bibinfo {author} {\bibfnamefont {H.}~\bibnamefont {Bernien}},
  \bibinfo {author} {\bibfnamefont {A.~E.}\ \bibnamefont {Dr{\'{e}}au}},
  \bibinfo {author} {\bibfnamefont {A.}~\bibnamefont {Reiserer}}, \bibinfo
  {author} {\bibfnamefont {N.}~\bibnamefont {Kalb}}, \bibinfo {author}
  {\bibfnamefont {M.~S.}\ \bibnamefont {Blok}}, \bibinfo {author}
  {\bibfnamefont {J.}~\bibnamefont {Ruitenberg}}, \bibinfo {author}
  {\bibfnamefont {R.~F.~L.}\ \bibnamefont {Vermeulen}}, \bibinfo {author}
  {\bibfnamefont {R.~N.}\ \bibnamefont {Schouten}}, \bibinfo {author}
  {\bibfnamefont {C.}~\bibnamefont {Abell{\'{a}}n}}, \bibinfo {author}
  {\bibfnamefont {W.}~\bibnamefont {Amaya}}, \bibinfo {author} {\bibfnamefont
  {V.}~\bibnamefont {Pruneri}}, \bibinfo {author} {\bibfnamefont {M.~W.}\
  \bibnamefont {Mitchell}}, \bibinfo {author} {\bibfnamefont {M.}~\bibnamefont
  {Markham}}, \bibinfo {author} {\bibfnamefont {D.~J.}\ \bibnamefont
  {Twitchen}}, \bibinfo {author} {\bibfnamefont {D.}~\bibnamefont {Elkouss}},
  \bibinfo {author} {\bibfnamefont {S.}~\bibnamefont {Wehner}}, \bibinfo
  {author} {\bibfnamefont {T.~H.}\ \bibnamefont {Taminiau}}, \ and\ \bibinfo
  {author} {\bibfnamefont {R.}~\bibnamefont {Hanson}},\ }\href {\doibase
  10.1038/nature15759} {\bibfield  {journal} {\bibinfo  {journal} {Nature}\
  }\textbf {\bibinfo {volume} {526}},\ \bibinfo {pages} {682} (\bibinfo {year}
  {2015})}\BibitemShut {NoStop}%
\bibitem [{\citenamefont {Wootters}(1998)}]{Wootters1998}%
  \BibitemOpen
  \bibfield  {author} {\bibinfo {author} {\bibfnamefont {W.~K.}\ \bibnamefont
  {Wootters}},\ }\href {\doibase 10.1098/rsta.1998.0244} {\bibfield  {journal}
  {\bibinfo  {journal} {Philosophical Transactions of the Royal Society A:
  Mathematical, Physical and Engineering Sciences}\ }\textbf {\bibinfo {volume}
  {356}},\ \bibinfo {pages} {1717} (\bibinfo {year} {1998})}\BibitemShut
  {NoStop}%
\bibitem [{\citenamefont {Kimble}(2008)}]{Kimble2008}%
  \BibitemOpen
  \bibfield  {author} {\bibinfo {author} {\bibfnamefont {H.~J.}\ \bibnamefont
  {Kimble}},\ }\href {\doibase 10.1038/nature07127} {\bibfield  {journal}
  {\bibinfo  {journal} {Nature}\ }\textbf {\bibinfo {volume} {453}},\ \bibinfo
  {pages} {1023} (\bibinfo {year} {2008})}\BibitemShut {NoStop}%
\bibitem [{\citenamefont {Lodahl}\ \emph {et~al.}(2015)\citenamefont {Lodahl},
  \citenamefont {Mahmoodian},\ and\ \citenamefont {Stobbe}}]{Lodahl2015}%
  \BibitemOpen
  \bibfield  {author} {\bibinfo {author} {\bibfnamefont {P.}~\bibnamefont
  {Lodahl}}, \bibinfo {author} {\bibfnamefont {S.}~\bibnamefont {Mahmoodian}},
  \ and\ \bibinfo {author} {\bibfnamefont {S.}~\bibnamefont {Stobbe}},\ }\href
  {\doibase 10.1103/RevModPhys.87.347} {\bibfield  {journal} {\bibinfo
  {journal} {Reviews of Modern Physics}\ }\textbf {\bibinfo {volume} {87}},\
  \bibinfo {pages} {347} (\bibinfo {year} {2015})}\BibitemShut {NoStop}%
\bibitem [{\citenamefont {Press}\ \emph {et~al.}(2008)\citenamefont {Press},
  \citenamefont {Ladd}, \citenamefont {Zhang},\ and\ \citenamefont
  {Yamamoto}}]{Press2008}%
  \BibitemOpen
  \bibfield  {author} {\bibinfo {author} {\bibfnamefont {D.}~\bibnamefont
  {Press}}, \bibinfo {author} {\bibfnamefont {T.~D.}\ \bibnamefont {Ladd}},
  \bibinfo {author} {\bibfnamefont {B.}~\bibnamefont {Zhang}}, \ and\ \bibinfo
  {author} {\bibfnamefont {Y.}~\bibnamefont {Yamamoto}},\ }\href {\doibase
  10.1038/nature07530} {\bibfield  {journal} {\bibinfo  {journal} {Nature}\
  }\textbf {\bibinfo {volume} {456}},\ \bibinfo {pages} {218} (\bibinfo {year}
  {2008})}\BibitemShut {NoStop}%
\bibitem [{\citenamefont {Brunner}\ \emph {et~al.}(2009)\citenamefont
  {Brunner}, \citenamefont {Gerardot}, \citenamefont {Dalgarno}, \citenamefont
  {W{\"{u}}st}, \citenamefont {Karrai}, \citenamefont {Stoltz}, \citenamefont
  {Petroff},\ and\ \citenamefont {Warburton}}]{Brunner2009}%
  \BibitemOpen
  \bibfield  {author} {\bibinfo {author} {\bibfnamefont {D.}~\bibnamefont
  {Brunner}}, \bibinfo {author} {\bibfnamefont {B.~D.}\ \bibnamefont
  {Gerardot}}, \bibinfo {author} {\bibfnamefont {P.~A.}\ \bibnamefont
  {Dalgarno}}, \bibinfo {author} {\bibfnamefont {G.}~\bibnamefont
  {W{\"{u}}st}}, \bibinfo {author} {\bibfnamefont {K.}~\bibnamefont {Karrai}},
  \bibinfo {author} {\bibfnamefont {N.~G.}\ \bibnamefont {Stoltz}}, \bibinfo
  {author} {\bibfnamefont {P.~M.}\ \bibnamefont {Petroff}}, \ and\ \bibinfo
  {author} {\bibfnamefont {R.~J.}\ \bibnamefont {Warburton}},\ }\href {\doibase
  10.1126/science.1173684} {\bibfield  {journal} {\bibinfo  {journal}
  {Science}\ }\textbf {\bibinfo {volume} {325}},\ \bibinfo {pages} {70}
  (\bibinfo {year} {2009})}\BibitemShut {NoStop}%
\bibitem [{\citenamefont {Schwartz}\ \emph {et~al.}(2015)\citenamefont
  {Schwartz}, \citenamefont {Schmidgall}, \citenamefont {Gantz}, \citenamefont
  {Cogan}, \citenamefont {Bordo}, \citenamefont {Don}, \citenamefont
  {Zielinski},\ and\ \citenamefont {Gershoni}}]{Schwartz2015}%
  \BibitemOpen
  \bibfield  {author} {\bibinfo {author} {\bibfnamefont {I.}~\bibnamefont
  {Schwartz}}, \bibinfo {author} {\bibfnamefont {E.~R.}\ \bibnamefont
  {Schmidgall}}, \bibinfo {author} {\bibfnamefont {L.}~\bibnamefont {Gantz}},
  \bibinfo {author} {\bibfnamefont {D.}~\bibnamefont {Cogan}}, \bibinfo
  {author} {\bibfnamefont {E.}~\bibnamefont {Bordo}}, \bibinfo {author}
  {\bibfnamefont {Y.}~\bibnamefont {Don}}, \bibinfo {author} {\bibfnamefont
  {M.}~\bibnamefont {Zielinski}}, \ and\ \bibinfo {author} {\bibfnamefont
  {D.}~\bibnamefont {Gershoni}},\ }\href {\doibase 10.1103/PhysRevX.5.011009}
  {\bibfield  {journal} {\bibinfo  {journal} {Physical Review X}\ }\textbf
  {\bibinfo {volume} {5}},\ \bibinfo {pages} {011009} (\bibinfo {year}
  {2015})}\BibitemShut {NoStop}%
\bibitem [{\citenamefont {Sun}\ \emph {et~al.}(2016)\citenamefont {Sun},
  \citenamefont {Delteil}, \citenamefont {Faelt},\ and\ \citenamefont
  {Imamoğlu}}]{Sun2016}%
  \BibitemOpen
  \bibfield  {author} {\bibinfo {author} {\bibfnamefont {Z.}~\bibnamefont
  {Sun}}, \bibinfo {author} {\bibfnamefont {A.}~\bibnamefont {Delteil}},
  \bibinfo {author} {\bibfnamefont {S.}~\bibnamefont {Faelt}}, \ and\ \bibinfo
  {author} {\bibfnamefont {A.}~\bibnamefont {Imamoğlu}},\ }\href {\doibase
  10.1103/PhysRevB.93.241302} {\bibfield  {journal} {\bibinfo  {journal}
  {Physical Review B}\ }\textbf {\bibinfo {volume} {93}},\ \bibinfo {pages}
  {241302} (\bibinfo {year} {2016})}\BibitemShut {NoStop}%
\bibitem [{\citenamefont {Schwartz}\ \emph {et~al.}(2016)\citenamefont
  {Schwartz}, \citenamefont {Cogan}, \citenamefont {Schmidgall}, \citenamefont
  {Don}, \citenamefont {Gantz}, \citenamefont {Kenneth}, \citenamefont
  {Lindner},\ and\ \citenamefont {Gershoni}}]{Schwartz2016}%
  \BibitemOpen
  \bibfield  {author} {\bibinfo {author} {\bibfnamefont {I.}~\bibnamefont
  {Schwartz}}, \bibinfo {author} {\bibfnamefont {D.}~\bibnamefont {Cogan}},
  \bibinfo {author} {\bibfnamefont {E.~R.}\ \bibnamefont {Schmidgall}},
  \bibinfo {author} {\bibfnamefont {Y.}~\bibnamefont {Don}}, \bibinfo {author}
  {\bibfnamefont {L.}~\bibnamefont {Gantz}}, \bibinfo {author} {\bibfnamefont
  {O.}~\bibnamefont {Kenneth}}, \bibinfo {author} {\bibfnamefont {N.~H.}\
  \bibnamefont {Lindner}}, \ and\ \bibinfo {author} {\bibfnamefont
  {D.}~\bibnamefont {Gershoni}},\ }\href {\doibase 10.1126/science.aah4758}
  {\bibfield  {journal} {\bibinfo  {journal} {Science}\ }\textbf {\bibinfo
  {volume} {49}},\ \bibinfo {pages} {1804} (\bibinfo {year}
  {2016})}\BibitemShut {NoStop}%
\bibitem [{\citenamefont {Press}\ \emph {et~al.}(2010)\citenamefont {Press},
  \citenamefont {{De Greve}}, \citenamefont {McMahon}, \citenamefont {Ladd},
  \citenamefont {Friess}, \citenamefont {Schneider}, \citenamefont {Kamp},
  \citenamefont {H{\"{o}}fling}, \citenamefont {Forchel},\ and\ \citenamefont
  {Yamamoto}}]{Press2010}%
  \BibitemOpen
  \bibfield  {author} {\bibinfo {author} {\bibfnamefont {D.}~\bibnamefont
  {Press}}, \bibinfo {author} {\bibfnamefont {K.}~\bibnamefont {{De Greve}}},
  \bibinfo {author} {\bibfnamefont {P.~L.}\ \bibnamefont {McMahon}}, \bibinfo
  {author} {\bibfnamefont {T.~D.}\ \bibnamefont {Ladd}}, \bibinfo {author}
  {\bibfnamefont {B.}~\bibnamefont {Friess}}, \bibinfo {author} {\bibfnamefont
  {C.}~\bibnamefont {Schneider}}, \bibinfo {author} {\bibfnamefont
  {M.}~\bibnamefont {Kamp}}, \bibinfo {author} {\bibfnamefont {S.}~\bibnamefont
  {H{\"{o}}fling}}, \bibinfo {author} {\bibfnamefont {A.}~\bibnamefont
  {Forchel}}, \ and\ \bibinfo {author} {\bibfnamefont {Y.}~\bibnamefont
  {Yamamoto}},\ }\href {\doibase 10.1038/nphoton.2010.83} {\bibfield  {journal}
  {\bibinfo  {journal} {Nature Photonics}\ }\textbf {\bibinfo {volume} {4}},\
  \bibinfo {pages} {367} (\bibinfo {year} {2010})}\BibitemShut {NoStop}%
\bibitem [{\citenamefont {Bechtold}\ \emph {et~al.}(2015)\citenamefont
  {Bechtold}, \citenamefont {Rauch}, \citenamefont {Li}, \citenamefont
  {Simmet}, \citenamefont {Ardelt}, \citenamefont {Regler}, \citenamefont
  {M{\"{u}}ller}, \citenamefont {Sinitsyn},\ and\ \citenamefont
  {Finley}}]{Bechtold2015}%
  \BibitemOpen
  \bibfield  {author} {\bibinfo {author} {\bibfnamefont {A.}~\bibnamefont
  {Bechtold}}, \bibinfo {author} {\bibfnamefont {D.}~\bibnamefont {Rauch}},
  \bibinfo {author} {\bibfnamefont {F.}~\bibnamefont {Li}}, \bibinfo {author}
  {\bibfnamefont {T.}~\bibnamefont {Simmet}}, \bibinfo {author} {\bibfnamefont
  {P.-L.}\ \bibnamefont {Ardelt}}, \bibinfo {author} {\bibfnamefont
  {A.}~\bibnamefont {Regler}}, \bibinfo {author} {\bibfnamefont
  {K.}~\bibnamefont {M{\"{u}}ller}}, \bibinfo {author} {\bibfnamefont {N.~A.}\
  \bibnamefont {Sinitsyn}}, \ and\ \bibinfo {author} {\bibfnamefont {J.~J.}\
  \bibnamefont {Finley}},\ }\href {\doibase 10.1038/nphys3470} {\bibfield
  {journal} {\bibinfo  {journal} {Nature Physics}\ }\textbf {\bibinfo {volume}
  {11}},\ \bibinfo {pages} {1005} (\bibinfo {year} {2015})}\BibitemShut
  {NoStop}%
\bibitem [{\citenamefont {Stockill}\ \emph {et~al.}(2016)\citenamefont
  {Stockill}, \citenamefont {{Le Gall}}, \citenamefont {Matthiesen},
  \citenamefont {Huthmacher}, \citenamefont {Clarke}, \citenamefont {Hugues},\
  and\ \citenamefont {Atat{\"{u}}re}}]{Stockill2016}%
  \BibitemOpen
  \bibfield  {author} {\bibinfo {author} {\bibfnamefont {R.}~\bibnamefont
  {Stockill}}, \bibinfo {author} {\bibfnamefont {C.}~\bibnamefont {{Le Gall}}},
  \bibinfo {author} {\bibfnamefont {C.}~\bibnamefont {Matthiesen}}, \bibinfo
  {author} {\bibfnamefont {L.}~\bibnamefont {Huthmacher}}, \bibinfo {author}
  {\bibfnamefont {E.}~\bibnamefont {Clarke}}, \bibinfo {author} {\bibfnamefont
  {M.}~\bibnamefont {Hugues}}, \ and\ \bibinfo {author} {\bibfnamefont
  {M.}~\bibnamefont {Atat{\"{u}}re}},\ }\href {\doibase 10.1038/ncomms12745}
  {\bibfield  {journal} {\bibinfo  {journal} {Nature Communications}\ }\textbf
  {\bibinfo {volume} {7}},\ \bibinfo {pages} {12745} (\bibinfo {year}
  {2016})}\BibitemShut {NoStop}%
\bibitem [{\citenamefont {Cabrillo}\ \emph {et~al.}(1998)\citenamefont
  {Cabrillo}, \citenamefont {Cirac}, \citenamefont {Garcia-Fernandez},\ and\
  \citenamefont {Zoller}}]{Cabrillo1998}%
  \BibitemOpen
  \bibfield  {author} {\bibinfo {author} {\bibfnamefont {C.}~\bibnamefont
  {Cabrillo}}, \bibinfo {author} {\bibfnamefont {J.~I.}\ \bibnamefont {Cirac}},
  \bibinfo {author} {\bibfnamefont {P.}~\bibnamefont {Garcia-Fernandez}}, \
  and\ \bibinfo {author} {\bibfnamefont {P.}~\bibnamefont {Zoller}},\ }\href
  {\doibase 10.1103/PhysRevA.59.1025} {\bibfield  {journal} {\bibinfo
  {journal} {Physical Review A}\ }\textbf {\bibinfo {volume} {59}},\ \bibinfo
  {pages} {10} (\bibinfo {year} {1998})}\BibitemShut {NoStop}%
\bibitem [{\citenamefont {Kuhlmann}\ \emph {et~al.}(2015)\citenamefont
  {Kuhlmann}, \citenamefont {Prechtel}, \citenamefont {Houel}, \citenamefont
  {Ludwig}, \citenamefont {Reuter}, \citenamefont {Wieck},\ and\ \citenamefont
  {Warburton}}]{Kuhlmann2015}%
  \BibitemOpen
  \bibfield  {author} {\bibinfo {author} {\bibfnamefont {A.~V.}\ \bibnamefont
  {Kuhlmann}}, \bibinfo {author} {\bibfnamefont {J.~H.}\ \bibnamefont
  {Prechtel}}, \bibinfo {author} {\bibfnamefont {J.}~\bibnamefont {Houel}},
  \bibinfo {author} {\bibfnamefont {A.}~\bibnamefont {Ludwig}}, \bibinfo
  {author} {\bibfnamefont {D.}~\bibnamefont {Reuter}}, \bibinfo {author}
  {\bibfnamefont {A.~D.}\ \bibnamefont {Wieck}}, \ and\ \bibinfo {author}
  {\bibfnamefont {R.~J.}\ \bibnamefont {Warburton}},\ }\href {\doibase
  10.1038/ncomms9204} {\bibfield  {journal} {\bibinfo  {journal} {Nature
  Communications}\ }\textbf {\bibinfo {volume} {6}},\ \bibinfo {pages} {8204}
  (\bibinfo {year} {2015})}\BibitemShut {NoStop}%
\bibitem [{\citenamefont {Slodi{\v{c}}ka}\ \emph {et~al.}(2013)\citenamefont
  {Slodi{\v{c}}ka}, \citenamefont {H{\'{e}}tet}, \citenamefont {R{\"{o}}ck},
  \citenamefont {Schindler}, \citenamefont {Hennrich},\ and\ \citenamefont
  {Blatt}}]{Slodicka2013}%
  \BibitemOpen
  \bibfield  {author} {\bibinfo {author} {\bibfnamefont {L.}~\bibnamefont
  {Slodi{\v{c}}ka}}, \bibinfo {author} {\bibfnamefont {G.}~\bibnamefont
  {H{\'{e}}tet}}, \bibinfo {author} {\bibfnamefont {N.}~\bibnamefont
  {R{\"{o}}ck}}, \bibinfo {author} {\bibfnamefont {P.}~\bibnamefont
  {Schindler}}, \bibinfo {author} {\bibfnamefont {M.}~\bibnamefont {Hennrich}},
  \ and\ \bibinfo {author} {\bibfnamefont {R.}~\bibnamefont {Blatt}},\ }\href
  {\doibase 10.1103/PhysRevLett.110.083603} {\bibfield  {journal} {\bibinfo
  {journal} {Physical Review Letters}\ }\textbf {\bibinfo {volume} {110}},\
  \bibinfo {pages} {083603} (\bibinfo {year} {2013})}\BibitemShut {NoStop}%
\bibitem [{\citenamefont {Luxmoore}\ \emph {et~al.}(2013)\citenamefont
  {Luxmoore}, \citenamefont {Wasley}, \citenamefont {Ramsay}, \citenamefont
  {Thijssen}, \citenamefont {Oulton}, \citenamefont {Hugues}, \citenamefont
  {Kasture}, \citenamefont {Achanta}, \citenamefont {Fox},\ and\ \citenamefont
  {Skolnick}}]{Luxmoore2013}%
  \BibitemOpen
  \bibfield  {author} {\bibinfo {author} {\bibfnamefont {I.~J.}\ \bibnamefont
  {Luxmoore}}, \bibinfo {author} {\bibfnamefont {N.~A.}\ \bibnamefont
  {Wasley}}, \bibinfo {author} {\bibfnamefont {A.~J.}\ \bibnamefont {Ramsay}},
  \bibinfo {author} {\bibfnamefont {A.~C.~T.}\ \bibnamefont {Thijssen}},
  \bibinfo {author} {\bibfnamefont {R.}~\bibnamefont {Oulton}}, \bibinfo
  {author} {\bibfnamefont {M.}~\bibnamefont {Hugues}}, \bibinfo {author}
  {\bibfnamefont {S.}~\bibnamefont {Kasture}}, \bibinfo {author} {\bibfnamefont
  {V.~G.}\ \bibnamefont {Achanta}}, \bibinfo {author} {\bibfnamefont {A.~M.}\
  \bibnamefont {Fox}}, \ and\ \bibinfo {author} {\bibfnamefont {M.~S.}\
  \bibnamefont {Skolnick}},\ }\href {\doibase 10.1103/PhysRevLett.110.037402}
  {\bibfield  {journal} {\bibinfo  {journal} {Physical Review Letters}\
  }\textbf {\bibinfo {volume} {110}},\ \bibinfo {pages} {037402} (\bibinfo
  {year} {2013})}\BibitemShut {NoStop}%
\bibitem [{\citenamefont {Gao}\ \emph {et~al.}(2012)\citenamefont {Gao},
  \citenamefont {Fallahi}, \citenamefont {Togan}, \citenamefont
  {Miguel-Sanchez},\ and\ \citenamefont {Imamoglu}}]{Gao2012}%
  \BibitemOpen
  \bibfield  {author} {\bibinfo {author} {\bibfnamefont {W.~B.}\ \bibnamefont
  {Gao}}, \bibinfo {author} {\bibfnamefont {P.}~\bibnamefont {Fallahi}},
  \bibinfo {author} {\bibfnamefont {E.}~\bibnamefont {Togan}}, \bibinfo
  {author} {\bibfnamefont {J.}~\bibnamefont {Miguel-Sanchez}}, \ and\ \bibinfo
  {author} {\bibfnamefont {A.}~\bibnamefont {Imamoglu}},\ }\href {\doibase
  10.1038/nature11573} {\bibfield  {journal} {\bibinfo  {journal} {Nature}\
  }\textbf {\bibinfo {volume} {491}},\ \bibinfo {pages} {426} (\bibinfo {year}
  {2012})}\BibitemShut {NoStop}%
\bibitem [{\citenamefont {{De Greve}}\ \emph {et~al.}(2012)\citenamefont {{De
  Greve}}, \citenamefont {Yu}, \citenamefont {McMahon}, \citenamefont {Pelc},
  \citenamefont {Natarajan}, \citenamefont {Kim}, \citenamefont {Abe},
  \citenamefont {Maier}, \citenamefont {Schneider}, \citenamefont {Kamp},
  \citenamefont {Hoefling}, \citenamefont {Hadfield}, \citenamefont {Forchel},
  \citenamefont {Fejer},\ and\ \citenamefont {Yamamoto}}]{Degreve2012}%
  \BibitemOpen
  \bibfield  {author} {\bibinfo {author} {\bibfnamefont {K.}~\bibnamefont {{De
  Greve}}}, \bibinfo {author} {\bibfnamefont {L.}~\bibnamefont {Yu}}, \bibinfo
  {author} {\bibfnamefont {P.~L.}\ \bibnamefont {McMahon}}, \bibinfo {author}
  {\bibfnamefont {J.~S.}\ \bibnamefont {Pelc}}, \bibinfo {author}
  {\bibfnamefont {C.~M.}\ \bibnamefont {Natarajan}}, \bibinfo {author}
  {\bibfnamefont {N.~Y.}\ \bibnamefont {Kim}}, \bibinfo {author} {\bibfnamefont
  {E.}~\bibnamefont {Abe}}, \bibinfo {author} {\bibfnamefont {S.}~\bibnamefont
  {Maier}}, \bibinfo {author} {\bibfnamefont {C.}~\bibnamefont {Schneider}},
  \bibinfo {author} {\bibfnamefont {M.}~\bibnamefont {Kamp}}, \bibinfo {author}
  {\bibfnamefont {S.}~\bibnamefont {Hoefling}}, \bibinfo {author}
  {\bibfnamefont {R.~H.}\ \bibnamefont {Hadfield}}, \bibinfo {author}
  {\bibfnamefont {A.}~\bibnamefont {Forchel}}, \bibinfo {author} {\bibfnamefont
  {M.~M.}\ \bibnamefont {Fejer}}, \ and\ \bibinfo {author} {\bibfnamefont
  {Y.}~\bibnamefont {Yamamoto}},\ }\href {\doibase 10.1038/nature11577}
  {\bibfield  {journal} {\bibinfo  {journal} {Nature}\ }\textbf {\bibinfo
  {volume} {491}},\ \bibinfo {pages} {421} (\bibinfo {year}
  {2012})}\BibitemShut {NoStop}%
\bibitem [{\citenamefont {Schaibley}\ \emph {et~al.}(2013)\citenamefont
  {Schaibley}, \citenamefont {Burgers}, \citenamefont {McCracken},
  \citenamefont {Duan}, \citenamefont {Berman}, \citenamefont {Steel},
  \citenamefont {Bracker}, \citenamefont {Gammon},\ and\ \citenamefont
  {Sham}}]{Schaibley2013}%
  \BibitemOpen
  \bibfield  {author} {\bibinfo {author} {\bibfnamefont {J.~R.}\ \bibnamefont
  {Schaibley}}, \bibinfo {author} {\bibfnamefont {A.~P.}\ \bibnamefont
  {Burgers}}, \bibinfo {author} {\bibfnamefont {G.~A.}\ \bibnamefont
  {McCracken}}, \bibinfo {author} {\bibfnamefont {L.-M.~M.}\ \bibnamefont
  {Duan}}, \bibinfo {author} {\bibfnamefont {P.~R.}\ \bibnamefont {Berman}},
  \bibinfo {author} {\bibfnamefont {D.~G.}\ \bibnamefont {Steel}}, \bibinfo
  {author} {\bibfnamefont {a.~S.}\ \bibnamefont {Bracker}}, \bibinfo {author}
  {\bibfnamefont {D.}~\bibnamefont {Gammon}}, \ and\ \bibinfo {author}
  {\bibfnamefont {L.~J.}\ \bibnamefont {Sham}},\ }\href {\doibase
  10.1103/PhysRevLett.110.167401} {\bibfield  {journal} {\bibinfo  {journal}
  {Physical Review Letters}\ }\textbf {\bibinfo {volume} {110}},\ \bibinfo
  {pages} {167401} (\bibinfo {year} {2013})}\BibitemShut {NoStop}%
\bibitem [{\citenamefont {Merkulov}\ \emph {et~al.}(2002)\citenamefont
  {Merkulov}, \citenamefont {Efros},\ and\ \citenamefont
  {Rosen}}]{Merkulov2002}%
  \BibitemOpen
  \bibfield  {author} {\bibinfo {author} {\bibfnamefont {I.~A.}\ \bibnamefont
  {Merkulov}}, \bibinfo {author} {\bibfnamefont {A.~L.}\ \bibnamefont {Efros}},
  \ and\ \bibinfo {author} {\bibfnamefont {M.}~\bibnamefont {Rosen}},\ }\href
  {\doibase 10.1103/PhysRevB.65.205309} {\bibfield  {journal} {\bibinfo
  {journal} {Physical Review B}\ }\textbf {\bibinfo {volume} {65}},\ \bibinfo
  {pages} {205309} (\bibinfo {year} {2002})}\BibitemShut {NoStop}%
\bibitem [{sup()}]{supplementary}%
  \BibitemOpen
  \href@noop {} {}\bibinfo {note} {Supplemental material available online.}\BibitemShut {Stop}%
\bibitem [{\citenamefont {Vittorini}\ \emph {et~al.}(2014)\citenamefont
  {Vittorini}, \citenamefont {Hucul}, \citenamefont {Inlek}, \citenamefont
  {Crocker},\ and\ \citenamefont {Monroe}}]{Vittorini2014}%
  \BibitemOpen
  \bibfield  {author} {\bibinfo {author} {\bibfnamefont {G.}~\bibnamefont
  {Vittorini}}, \bibinfo {author} {\bibfnamefont {D.}~\bibnamefont {Hucul}},
  \bibinfo {author} {\bibfnamefont {I.~V.}\ \bibnamefont {Inlek}}, \bibinfo
  {author} {\bibfnamefont {C.}~\bibnamefont {Crocker}}, \ and\ \bibinfo
  {author} {\bibfnamefont {C.}~\bibnamefont {Monroe}},\ }\href {\doibase
  10.1103/PhysRevA.90.040302} {\bibfield  {journal} {\bibinfo  {journal}
  {Physical Review A}\ }\textbf {\bibinfo {volume} {90}},\ \bibinfo {pages}
  {040302} (\bibinfo {year} {2014})}\BibitemShut {NoStop}%
\bibitem [{\citenamefont {Hong}\ \emph {et~al.}(1987)\citenamefont {Hong},
  \citenamefont {Ou},\ and\ \citenamefont {Mandel}}]{Hong1987}%
  \BibitemOpen
  \bibfield  {author} {\bibinfo {author} {\bibfnamefont {C.~K.}\ \bibnamefont
  {Hong}}, \bibinfo {author} {\bibfnamefont {Z.~Y.}\ \bibnamefont {Ou}}, \ and\
  \bibinfo {author} {\bibfnamefont {L.}~\bibnamefont {Mandel}},\ }\href
  {\doibase 10.1103/PhysRevLett.59.2044} {\bibfield  {journal} {\bibinfo
  {journal} {Physical Review Letters}\ }\textbf {\bibinfo {volume} {59}},\
  \bibinfo {pages} {2044} (\bibinfo {year} {1987})}\BibitemShut {NoStop}%
\bibitem [{\citenamefont {Santori}\ \emph {et~al.}(2009)\citenamefont
  {Santori}, \citenamefont {Fattal}, \citenamefont {Fu}, \citenamefont
  {Barclay},\ and\ \citenamefont {Beausoleil}}]{Santori2009}%
  \BibitemOpen
  \bibfield  {author} {\bibinfo {author} {\bibfnamefont {C.}~\bibnamefont
  {Santori}}, \bibinfo {author} {\bibfnamefont {D.}~\bibnamefont {Fattal}},
  \bibinfo {author} {\bibfnamefont {K.-M. M.~C.}\ \bibnamefont {Fu}}, \bibinfo
  {author} {\bibfnamefont {P.~E.}\ \bibnamefont {Barclay}}, \ and\ \bibinfo
  {author} {\bibfnamefont {R.~G.}\ \bibnamefont {Beausoleil}},\ }\href
  {\doibase http://dx.doi.org/10.1088/1367-2630/11/12/123009} {\bibfield
  {journal} {\bibinfo  {journal} {New Journal of Physics}\ }\textbf {\bibinfo
  {volume} {11}},\ \bibinfo {pages} {123009} (\bibinfo {year}
  {2009})}\BibitemShut {NoStop}%
\bibitem [{\citenamefont {Kuhlmann}\ \emph {et~al.}(2013)\citenamefont
  {Kuhlmann}, \citenamefont {Houel}, \citenamefont {Ludwig}, \citenamefont
  {Greuter}, \citenamefont {Reuter}, \citenamefont {Wieck}, \citenamefont
  {Poggio},\ and\ \citenamefont {Warburton}}]{Kuhlmann2013}%
  \BibitemOpen
  \bibfield  {author} {\bibinfo {author} {\bibfnamefont {A.~V.}\ \bibnamefont
  {Kuhlmann}}, \bibinfo {author} {\bibfnamefont {J.}~\bibnamefont {Houel}},
  \bibinfo {author} {\bibfnamefont {A.}~\bibnamefont {Ludwig}}, \bibinfo
  {author} {\bibfnamefont {L.}~\bibnamefont {Greuter}}, \bibinfo {author}
  {\bibfnamefont {D.}~\bibnamefont {Reuter}}, \bibinfo {author} {\bibfnamefont
  {A.~D.}\ \bibnamefont {Wieck}}, \bibinfo {author} {\bibfnamefont
  {M.}~\bibnamefont {Poggio}}, \ and\ \bibinfo {author} {\bibfnamefont {R.~J.}\
  \bibnamefont {Warburton}},\ }\href {\doibase 10.1038/nphys2688} {\bibfield
  {journal} {\bibinfo  {journal} {Nature Physics}\ }\textbf {\bibinfo {volume}
  {9}},\ \bibinfo {pages} {570} (\bibinfo {year} {2013})}\BibitemShut {NoStop}%
\bibitem [{\citenamefont {Matthiesen}\ \emph {et~al.}(2014)\citenamefont
  {Matthiesen}, \citenamefont {Stanley}, \citenamefont {Hugues}, \citenamefont
  {Clarke},\ and\ \citenamefont {Atat{\"{u}}re}}]{Matthiesen2014}%
  \BibitemOpen
  \bibfield  {author} {\bibinfo {author} {\bibfnamefont {C.}~\bibnamefont
  {Matthiesen}}, \bibinfo {author} {\bibfnamefont {M.~J.}\ \bibnamefont
  {Stanley}}, \bibinfo {author} {\bibfnamefont {M.}~\bibnamefont {Hugues}},
  \bibinfo {author} {\bibfnamefont {E.}~\bibnamefont {Clarke}}, \ and\ \bibinfo
  {author} {\bibfnamefont {M.}~\bibnamefont {Atat{\"{u}}re}},\ }\href {\doibase
  10.1038/srep04911} {\bibfield  {journal} {\bibinfo  {journal} {Scientific
  Reports}\ }\textbf {\bibinfo {volume} {4}},\ \bibinfo {pages} {4911}
  (\bibinfo {year} {2014})}\BibitemShut {NoStop}%
\bibitem [{\citenamefont {Moehring}\ \emph {et~al.}(2007)\citenamefont
  {Moehring}, \citenamefont {Maunz}, \citenamefont {Olmschenk}, \citenamefont
  {Younge}, \citenamefont {Matsukevich}, \citenamefont {Duan},\ and\
  \citenamefont {Monroe}}]{Moehring2007}%
  \BibitemOpen
  \bibfield  {author} {\bibinfo {author} {\bibfnamefont {D.~L.}\ \bibnamefont
  {Moehring}}, \bibinfo {author} {\bibfnamefont {P.}~\bibnamefont {Maunz}},
  \bibinfo {author} {\bibfnamefont {S.}~\bibnamefont {Olmschenk}}, \bibinfo
  {author} {\bibfnamefont {K.~C.}\ \bibnamefont {Younge}}, \bibinfo {author}
  {\bibfnamefont {D.~N.}\ \bibnamefont {Matsukevich}}, \bibinfo {author}
  {\bibfnamefont {L.-M.}\ \bibnamefont {Duan}}, \ and\ \bibinfo {author}
  {\bibfnamefont {C.}~\bibnamefont {Monroe}},\ }\href {\doibase
  10.1038/nature06118} {\bibfield  {journal} {\bibinfo  {journal} {Nature}\
  }\textbf {\bibinfo {volume} {449}},\ \bibinfo {pages} {68} (\bibinfo {year}
  {2007})}\BibitemShut {NoStop}%
\bibitem [{\citenamefont {Maunz}\ \emph {et~al.}(2009)\citenamefont {Maunz},
  \citenamefont {Olmschenk}, \citenamefont {Hayes}, \citenamefont
  {Matsukevich}, \citenamefont {Duan},\ and\ \citenamefont
  {Monroe}}]{Maunz2009}%
  \BibitemOpen
  \bibfield  {author} {\bibinfo {author} {\bibfnamefont {P.}~\bibnamefont
  {Maunz}}, \bibinfo {author} {\bibfnamefont {S.}~\bibnamefont {Olmschenk}},
  \bibinfo {author} {\bibfnamefont {D.}~\bibnamefont {Hayes}}, \bibinfo
  {author} {\bibfnamefont {D.~N.}\ \bibnamefont {Matsukevich}}, \bibinfo
  {author} {\bibfnamefont {L.-M.}\ \bibnamefont {Duan}}, \ and\ \bibinfo
  {author} {\bibfnamefont {C.}~\bibnamefont {Monroe}},\ }\href {\doibase
  10.1103/PhysRevLett.102.250502} {\bibfield  {journal} {\bibinfo  {journal}
  {Physical Review Letters}\ }\textbf {\bibinfo {volume} {102}},\ \bibinfo
  {pages} {250502} (\bibinfo {year} {2009})}\BibitemShut {NoStop}%
\bibitem [{\citenamefont {Hofmann}\ \emph {et~al.}(2012)\citenamefont
  {Hofmann}, \citenamefont {Krug}, \citenamefont {Ortegel}, \citenamefont
  {Gerard}, \citenamefont {Weber}, \citenamefont {Rosenfeld},\ and\
  \citenamefont {Weinfurter}}]{Hofmann2012}%
  \BibitemOpen
  \bibfield  {author} {\bibinfo {author} {\bibfnamefont {J.}~\bibnamefont
  {Hofmann}}, \bibinfo {author} {\bibfnamefont {M.}~\bibnamefont {Krug}},
  \bibinfo {author} {\bibfnamefont {N.}~\bibnamefont {Ortegel}}, \bibinfo
  {author} {\bibfnamefont {L.}~\bibnamefont {Gerard}}, \bibinfo {author}
  {\bibfnamefont {M.}~\bibnamefont {Weber}}, \bibinfo {author} {\bibfnamefont
  {W.}~\bibnamefont {Rosenfeld}}, \ and\ \bibinfo {author} {\bibfnamefont
  {H.}~\bibnamefont {Weinfurter}},\ }\href {\doibase 10.1126/science.1221856}
  {\bibfield  {journal} {\bibinfo  {journal} {Science}\ }\textbf {\bibinfo
  {volume} {337}},\ \bibinfo {pages} {72} (\bibinfo {year} {2012})}\BibitemShut
  {NoStop}%
\bibitem [{\citenamefont {Hucul}\ \emph {et~al.}(2014)\citenamefont {Hucul},
  \citenamefont {Inlek}, \citenamefont {Vittorini}, \citenamefont {Crocker},
  \citenamefont {Debnath}, \citenamefont {Clark},\ and\ \citenamefont
  {Monroe}}]{Hucul2014}%
  \BibitemOpen
  \bibfield  {author} {\bibinfo {author} {\bibfnamefont {D.}~\bibnamefont
  {Hucul}}, \bibinfo {author} {\bibfnamefont {I.~V.}\ \bibnamefont {Inlek}},
  \bibinfo {author} {\bibfnamefont {G.}~\bibnamefont {Vittorini}}, \bibinfo
  {author} {\bibfnamefont {C.}~\bibnamefont {Crocker}}, \bibinfo {author}
  {\bibfnamefont {S.}~\bibnamefont {Debnath}}, \bibinfo {author} {\bibfnamefont
  {S.~M.}\ \bibnamefont {Clark}}, \ and\ \bibinfo {author} {\bibfnamefont
  {C.}~\bibnamefont {Monroe}},\ }\href {\doibase 10.1038/nphys3150} {\bibfield
  {journal} {\bibinfo  {journal} {Nature Physics}\ }\textbf {\bibinfo {volume}
  {11}},\ \bibinfo {pages} {37} (\bibinfo {year} {2014})}\BibitemShut {NoStop}%
\bibitem [{\citenamefont {Bernien}\ \emph {et~al.}(2013)\citenamefont
  {Bernien}, \citenamefont {Hensen}, \citenamefont {Pfaff}, \citenamefont
  {Koolstra}, \citenamefont {Blok}, \citenamefont {Robledo}, \citenamefont
  {Taminiau}, \citenamefont {Markham}, \citenamefont {Twitchen}, \citenamefont
  {Childress},\ and\ \citenamefont {Hanson}}]{Bernien2013}%
  \BibitemOpen
  \bibfield  {author} {\bibinfo {author} {\bibfnamefont {H.}~\bibnamefont
  {Bernien}}, \bibinfo {author} {\bibfnamefont {B.}~\bibnamefont {Hensen}},
  \bibinfo {author} {\bibfnamefont {W.}~\bibnamefont {Pfaff}}, \bibinfo
  {author} {\bibfnamefont {G.}~\bibnamefont {Koolstra}}, \bibinfo {author}
  {\bibfnamefont {M.~S.}\ \bibnamefont {Blok}}, \bibinfo {author}
  {\bibfnamefont {L.}~\bibnamefont {Robledo}}, \bibinfo {author} {\bibfnamefont
  {T.~H.}\ \bibnamefont {Taminiau}}, \bibinfo {author} {\bibfnamefont
  {M.}~\bibnamefont {Markham}}, \bibinfo {author} {\bibfnamefont {D.~J.}\
  \bibnamefont {Twitchen}}, \bibinfo {author} {\bibfnamefont {L.}~\bibnamefont
  {Childress}}, \ and\ \bibinfo {author} {\bibfnamefont {R.}~\bibnamefont
  {Hanson}},\ }\href {\doibase 10.1038/nature12016
  http://www.nature.com/nature/journal/vaop/ncurrent/abs/nature12016.html#supplementary-information}
  {\bibfield  {journal} {\bibinfo  {journal} {Nature}\ }\textbf {\bibinfo
  {volume} {497}},\ \bibinfo {pages} {86} (\bibinfo {year} {2013})}\BibitemShut
  {NoStop}%
\bibitem [{\citenamefont {Pfaff}\ \emph {et~al.}(2014)\citenamefont {Pfaff},
  \citenamefont {Hensen}, \citenamefont {Bernien}, \citenamefont {van Dam},
  \citenamefont {Blok}, \citenamefont {Taminiau}, \citenamefont {Tiggelman},
  \citenamefont {Schouten}, \citenamefont {Markham}, \citenamefont {Twitchen},\
  and\ \citenamefont {Hanson}}]{Pfaff2014}%
  \BibitemOpen
  \bibfield  {author} {\bibinfo {author} {\bibfnamefont {W.}~\bibnamefont
  {Pfaff}}, \bibinfo {author} {\bibfnamefont {B.~J.}\ \bibnamefont {Hensen}},
  \bibinfo {author} {\bibfnamefont {H.}~\bibnamefont {Bernien}}, \bibinfo
  {author} {\bibfnamefont {S.~B.}\ \bibnamefont {van Dam}}, \bibinfo {author}
  {\bibfnamefont {M.~S.}\ \bibnamefont {Blok}}, \bibinfo {author}
  {\bibfnamefont {T.~H.}\ \bibnamefont {Taminiau}}, \bibinfo {author}
  {\bibfnamefont {M.~J.}\ \bibnamefont {Tiggelman}}, \bibinfo {author}
  {\bibfnamefont {R.~N.}\ \bibnamefont {Schouten}}, \bibinfo {author}
  {\bibfnamefont {M.}~\bibnamefont {Markham}}, \bibinfo {author} {\bibfnamefont
  {D.~J.}\ \bibnamefont {Twitchen}}, \ and\ \bibinfo {author} {\bibfnamefont
  {R.}~\bibnamefont {Hanson}},\ }\href {\doibase 10.1126/science.1253512}
  {\bibfield  {journal} {\bibinfo  {journal} {Science}\ }\textbf {\bibinfo
  {volume} {345}},\ \bibinfo {pages} {532} (\bibinfo {year}
  {2014})}\BibitemShut {NoStop}%
\bibitem [{\citenamefont {Narla}\ \emph {et~al.}(2016)\citenamefont {Narla},
  \citenamefont {Shankar}, \citenamefont {Hatridge}, \citenamefont {Leghtas},
  \citenamefont {Sliwa}, \citenamefont {Zalys-Geller}, \citenamefont
  {Mundhada}, \citenamefont {Pfaff}, \citenamefont {Frunzio}, \citenamefont
  {Schoelkopf},\ and\ \citenamefont {Devoret}}]{Narla2016}%
  \BibitemOpen
  \bibfield  {author} {\bibinfo {author} {\bibfnamefont {A.}~\bibnamefont
  {Narla}}, \bibinfo {author} {\bibfnamefont {S.}~\bibnamefont {Shankar}},
  \bibinfo {author} {\bibfnamefont {M.}~\bibnamefont {Hatridge}}, \bibinfo
  {author} {\bibfnamefont {Z.}~\bibnamefont {Leghtas}}, \bibinfo {author}
  {\bibfnamefont {K.~M.}\ \bibnamefont {Sliwa}}, \bibinfo {author}
  {\bibfnamefont {E.}~\bibnamefont {Zalys-Geller}}, \bibinfo {author}
  {\bibfnamefont {S.~O.}\ \bibnamefont {Mundhada}}, \bibinfo {author}
  {\bibfnamefont {W.}~\bibnamefont {Pfaff}}, \bibinfo {author} {\bibfnamefont
  {L.}~\bibnamefont {Frunzio}}, \bibinfo {author} {\bibfnamefont {R.~J.}\
  \bibnamefont {Schoelkopf}}, \ and\ \bibinfo {author} {\bibfnamefont {M.~H.}\
  \bibnamefont {Devoret}},\ }\href {\doibase 10.1103/PhysRevX.6.031036}
  {\bibfield  {journal} {\bibinfo  {journal} {Physical Review X}\ }\textbf
  {\bibinfo {volume} {6}},\ \bibinfo {pages} {031036} (\bibinfo {year}
  {2016})}\BibitemShut {NoStop}%
\bibitem [{\citenamefont {Delteil}\ \emph {et~al.}(2015)\citenamefont
  {Delteil}, \citenamefont {Sun}, \citenamefont {Gao}, \citenamefont {Togan},
  \citenamefont {Faelt},\ and\ \citenamefont {Imamoglu}}]{Delteil2015}%
  \BibitemOpen
  \bibfield  {author} {\bibinfo {author} {\bibfnamefont {A.}~\bibnamefont
  {Delteil}}, \bibinfo {author} {\bibfnamefont {Z.}~\bibnamefont {Sun}},
  \bibinfo {author} {\bibfnamefont {W.-b.}\ \bibnamefont {Gao}}, \bibinfo
  {author} {\bibfnamefont {E.}~\bibnamefont {Togan}}, \bibinfo {author}
  {\bibfnamefont {S.}~\bibnamefont {Faelt}}, \ and\ \bibinfo {author}
  {\bibfnamefont {A.}~\bibnamefont {Imamoglu}},\ }\href {\doibase
  10.1038/nphys3605} {\bibfield  {journal} {\bibinfo  {journal} {Nature
  Physics}\ }\textbf {\bibinfo {volume} {12}},\ \bibinfo {pages} {218}
  (\bibinfo {year} {2015})}\BibitemShut {NoStop}%
\bibitem [{\citenamefont {Somaschi}\ \emph {et~al.}(2016)\citenamefont
  {Somaschi}, \citenamefont {Giesz}, \citenamefont {{De Santis}}, \citenamefont
  {Loredo}, \citenamefont {Almeida}, \citenamefont {Hornecker}, \citenamefont
  {Portalupi}, \citenamefont {Grange}, \citenamefont {Ant{\'{o}}n},
  \citenamefont {Demory}, \citenamefont {G{\'{o}}mez}, \citenamefont {Sagnes},
  \citenamefont {Lanzillotti-Kimura}, \citenamefont {Lema{\'{i}}tre},
  \citenamefont {Auffeves}, \citenamefont {White}, \citenamefont {Lanco},\ and\
  \citenamefont {Senellart}}]{Somaschi2015}%
  \BibitemOpen
  \bibfield  {author} {\bibinfo {author} {\bibfnamefont {N.}~\bibnamefont
  {Somaschi}}, \bibinfo {author} {\bibfnamefont {V.}~\bibnamefont {Giesz}},
  \bibinfo {author} {\bibfnamefont {L.}~\bibnamefont {{De Santis}}}, \bibinfo
  {author} {\bibfnamefont {J.~C.}\ \bibnamefont {Loredo}}, \bibinfo {author}
  {\bibfnamefont {M.~P.}\ \bibnamefont {Almeida}}, \bibinfo {author}
  {\bibfnamefont {G.}~\bibnamefont {Hornecker}}, \bibinfo {author}
  {\bibfnamefont {S.~L.}\ \bibnamefont {Portalupi}}, \bibinfo {author}
  {\bibfnamefont {T.}~\bibnamefont {Grange}}, \bibinfo {author} {\bibfnamefont
  {C.}~\bibnamefont {Ant{\'{o}}n}}, \bibinfo {author} {\bibfnamefont
  {J.}~\bibnamefont {Demory}}, \bibinfo {author} {\bibfnamefont
  {C.}~\bibnamefont {G{\'{o}}mez}}, \bibinfo {author} {\bibfnamefont
  {I.}~\bibnamefont {Sagnes}}, \bibinfo {author} {\bibfnamefont {N.~D.}\
  \bibnamefont {Lanzillotti-Kimura}}, \bibinfo {author} {\bibfnamefont
  {A.}~\bibnamefont {Lema{\'{i}}tre}}, \bibinfo {author} {\bibfnamefont
  {A.}~\bibnamefont {Auffeves}}, \bibinfo {author} {\bibfnamefont {A.~G.}\
  \bibnamefont {White}}, \bibinfo {author} {\bibfnamefont {L.}~\bibnamefont
  {Lanco}}, \ and\ \bibinfo {author} {\bibfnamefont {P.}~\bibnamefont
  {Senellart}},\ }\href {\doibase 10.1038/nphoton.2016.23} {\bibfield
  {journal} {\bibinfo  {journal} {Nature Photonics}\ }\textbf {\bibinfo
  {volume} {10}},\ \bibinfo {pages} {340} (\bibinfo {year} {2016})}\BibitemShut
  {NoStop}%
\bibitem [{\citenamefont {Wang}\ \emph {et~al.}(2016)\citenamefont {Wang},
  \citenamefont {Duan}, \citenamefont {Li}, \citenamefont {Chen}, \citenamefont
  {Li}, \citenamefont {He}, \citenamefont {Chen}, \citenamefont {He},
  \citenamefont {Ding}, \citenamefont {Peng}, \citenamefont {Schneider},
  \citenamefont {Kamp}, \citenamefont {H{\"{o}}fling}, \citenamefont {Lu},\
  and\ \citenamefont {Pan}}]{Wang2016}%
  \BibitemOpen
  \bibfield  {author} {\bibinfo {author} {\bibfnamefont {H.}~\bibnamefont
  {Wang}}, \bibinfo {author} {\bibfnamefont {Z.-C.}\ \bibnamefont {Duan}},
  \bibinfo {author} {\bibfnamefont {Y.-H.}\ \bibnamefont {Li}}, \bibinfo
  {author} {\bibfnamefont {S.}~\bibnamefont {Chen}}, \bibinfo {author}
  {\bibfnamefont {J.-P.}\ \bibnamefont {Li}}, \bibinfo {author} {\bibfnamefont
  {Y.-M.}\ \bibnamefont {He}}, \bibinfo {author} {\bibfnamefont {M.-C.}\
  \bibnamefont {Chen}}, \bibinfo {author} {\bibfnamefont {Y.}~\bibnamefont
  {He}}, \bibinfo {author} {\bibfnamefont {X.}~\bibnamefont {Ding}}, \bibinfo
  {author} {\bibfnamefont {C.-Z.}\ \bibnamefont {Peng}}, \bibinfo {author}
  {\bibfnamefont {C.}~\bibnamefont {Schneider}}, \bibinfo {author}
  {\bibfnamefont {M.}~\bibnamefont {Kamp}}, \bibinfo {author} {\bibfnamefont
  {S.}~\bibnamefont {H{\"{o}}fling}}, \bibinfo {author} {\bibfnamefont {C.-Y.}\
  \bibnamefont {Lu}}, \ and\ \bibinfo {author} {\bibfnamefont {J.-W.}\
  \bibnamefont {Pan}},\ }\href {\doibase 10.1103/PhysRevLett.116.213601}
  {\bibfield  {journal} {\bibinfo  {journal} {Physical Review Letters}\
  }\textbf {\bibinfo {volume} {116}},\ \bibinfo {pages} {213601} (\bibinfo
  {year} {2016})}\BibitemShut {NoStop}%
\bibitem [{\citenamefont {Prechtel}\ \emph {et~al.}(2016)\citenamefont
  {Prechtel}, \citenamefont {Kuhlmann}, \citenamefont {Houel}, \citenamefont
  {Ludwig}, \citenamefont {Valentin}, \citenamefont {Wieck},\ and\
  \citenamefont {Warburton}}]{Prechtel2016}%
  \BibitemOpen
  \bibfield  {author} {\bibinfo {author} {\bibfnamefont {J.~H.}\ \bibnamefont
  {Prechtel}}, \bibinfo {author} {\bibfnamefont {A.~V.}\ \bibnamefont
  {Kuhlmann}}, \bibinfo {author} {\bibfnamefont {J.}~\bibnamefont {Houel}},
  \bibinfo {author} {\bibfnamefont {A.}~\bibnamefont {Ludwig}}, \bibinfo
  {author} {\bibfnamefont {S.~R.}\ \bibnamefont {Valentin}}, \bibinfo {author}
  {\bibfnamefont {A.~D.}\ \bibnamefont {Wieck}}, \ and\ \bibinfo {author}
  {\bibfnamefont {R.~J.}\ \bibnamefont {Warburton}},\ }\href {\doibase
  10.1038/nmat4704} {\bibfield  {journal} {\bibinfo  {journal} {Nature
  Materials}\ }\textbf {\bibinfo {volume} {15}},\ \bibinfo {pages} {981}
  (\bibinfo {year} {2016})}\BibitemShut {NoStop}%
\bibitem [{\citenamefont {Monroe}\ \emph {et~al.}(2014)\citenamefont {Monroe},
  \citenamefont {Raussendorf}, \citenamefont {Ruthven}, \citenamefont {Brown},
  \citenamefont {Maunz}, \citenamefont {Duan},\ and\ \citenamefont
  {Kim}}]{Monroe2014}%
  \BibitemOpen
  \bibfield  {author} {\bibinfo {author} {\bibfnamefont {C.}~\bibnamefont
  {Monroe}}, \bibinfo {author} {\bibfnamefont {R.}~\bibnamefont {Raussendorf}},
  \bibinfo {author} {\bibfnamefont {A.}~\bibnamefont {Ruthven}}, \bibinfo
  {author} {\bibfnamefont {K.~R.}\ \bibnamefont {Brown}}, \bibinfo {author}
  {\bibfnamefont {P.}~\bibnamefont {Maunz}}, \bibinfo {author} {\bibfnamefont
  {L.-M.}\ \bibnamefont {Duan}}, \ and\ \bibinfo {author} {\bibfnamefont
  {J.}~\bibnamefont {Kim}},\ }\href {\doibase 10.1103/PhysRevA.89.022317}
  {\bibfield  {journal} {\bibinfo  {journal} {Physical Review A}\ }\textbf
  {\bibinfo {volume} {89}},\ \bibinfo {pages} {022317} (\bibinfo {year}
  {2014})}\BibitemShut {NoStop}%
\bibitem [{\citenamefont {Kim}\ \emph {et~al.}(2011)\citenamefont {Kim},
  \citenamefont {Carter}, \citenamefont {Greilich}, \citenamefont {Bracker},\
  and\ \citenamefont {Gammon}}]{Kim2011}%
  \BibitemOpen
  \bibfield  {author} {\bibinfo {author} {\bibfnamefont {D.}~\bibnamefont
  {Kim}}, \bibinfo {author} {\bibfnamefont {S.~G.}\ \bibnamefont {Carter}},
  \bibinfo {author} {\bibfnamefont {A.}~\bibnamefont {Greilich}}, \bibinfo
  {author} {\bibfnamefont {A.}~\bibnamefont {Bracker}}, \ and\ \bibinfo
  {author} {\bibfnamefont {D.}~\bibnamefont {Gammon}},\ }\href {\doibase
  10.1038/nphys1863} {\bibfield  {journal} {\bibinfo  {journal} {Nature
  Physics}\ }\textbf {\bibinfo {volume} {7}},\ \bibinfo {pages} {24} (\bibinfo
  {year} {2011})}\BibitemShut {NoStop}%
\bibitem [{\citenamefont {Kim}\ \emph {et~al.}(2016)\citenamefont {Kim},
  \citenamefont {Kiselev}, \citenamefont {Ross}, \citenamefont {Rakher},
  \citenamefont {Jones},\ and\ \citenamefont {Ladd}}]{Kim2016}%
  \BibitemOpen
  \bibfield  {author} {\bibinfo {author} {\bibfnamefont {D.}~\bibnamefont
  {Kim}}, \bibinfo {author} {\bibfnamefont {A.~A.}\ \bibnamefont {Kiselev}},
  \bibinfo {author} {\bibfnamefont {R.~S.}\ \bibnamefont {Ross}}, \bibinfo
  {author} {\bibfnamefont {M.~T.}\ \bibnamefont {Rakher}}, \bibinfo {author}
  {\bibfnamefont {C.}~\bibnamefont {Jones}}, \ and\ \bibinfo {author}
  {\bibfnamefont {T.~D.}\ \bibnamefont {Ladd}},\ }\href {\doibase
  10.1103/PhysRevApplied.5.024014} {\bibfield  {journal} {\bibinfo  {journal}
  {Physical Review Applied}\ }\textbf {\bibinfo {volume} {5}},\ \bibinfo
  {pages} {024014} (\bibinfo {year} {2016})}\BibitemShut {NoStop}%
\end{thebibliography}
\end{document}